\documentclass[12pt,preprint]{aastex}
\usepackage{graphics,graphicx}
\usepackage{epsfig}
\usepackage{amsmath}

\begin{document}
\def\etal {{\it et al. }}

\title{The Three-Dimensional Structure of Interior Ejecta in Cassiopeia A at High Spectral Resolution}
\author{Karl Isensee\altaffilmark{1}, Lawrence Rudnick\altaffilmark{1}, Tracey DeLaney\altaffilmark{2}, J.D. Smith\altaffilmark{3}, Jeonghee Rho\altaffilmark{4}, William T. Reach\altaffilmark{4}, Takashi Kozasa\altaffilmark{5}, and Haley Gomez\altaffilmark{6}}

\altaffiltext{1}{Astronomy Department, University of Minnesota, Minneapolis,
MN 55455; isensee@astro.umn.edu, larry@astro.umn.edu}
\altaffiltext{2}{Massachusetts Institute of Technology, Kavli Institute for
Astrophysics and Space Research, 77 Massachusetts Avenue, NE80, Cambridge, MA
02139; tdelaney@space.mit.edu}
\altaffiltext{3}{Ritter Astrophysical Observatory, University of Toledo,
Toledo, OH 43606 ; jd.smith@utoledo.edu}
\altaffiltext{4}{\emph{Spitzer} Science Center, California Institute of
Technology, MS 220-6, Pasadena, CA 91125; rho@ipac.caltech.edu,
reach@ipac.caltech.edu}
\altaffiltext{5}{Department of Cosmosciences, Graduate School of Science,
Hokkaido University, Sapporo 060-0810, Japan; kozasa@mail.sci.hokudai.ac.jp}
\altaffiltext{6}{School of Physics and Astronomy, University of Wales, P.O.
Box 913, Cardiff CF24 3YB, UK; haley.morgan@astro.cf.ac.uk}

\begin{abstract} We used the \emph{Spitzer Space Telescope}'s Infrared Spectrograph to create a high resolution spectral map of the central region of the Cassiopeia A supernova remnant, allowing us to make a Doppler reconstruction of its 3D structure.  The ejecta responsible for this emission have not yet encountered the remnant's reverse shock or the circumstellar medium, making it an ideal laboratory for exploring the dynamics of the supernova explosion itself.  We observe that the O, Si, and S ejecta can form both sheet-like structures as well as filaments.  Si and O, which come from different nucleosynthetic layers of the star, are observed to be coincident in velocity space in some regions, and separated by 500 km~s$^{-1}$ or more in others.  Ejecta traveling toward us are, on average,  $\sim$900 km~s$^{-1}$ slower than the material traveling away from us.  We compare our observations to recent supernova explosion models and find that no single model can simultaneously reproduce all the observed features.  However, models of different supernova explosions can collectively produce the observed geometries and structures of the interior emission.  We use the results from the models to address the conditions during the supernova explosion, concentrating on asymmetries in the shock structure.  We also predict that the back surface of Cassiopeia A will begin brightening in $\sim$30 years, and the front surface in $\sim$100 years.

\end{abstract}


\section{Introduction}

Cassiopeia A (Cas~A) is the second youngest known supernova remnant (SNR) in our galaxy, with only the recently discovered G1.9+0.3 being younger \citep{rey08}.  Extensive observations in the radio, infrared, optical, and X-ray give an estimated explosion date of around 1680 AD \citep{thor01, fesen06}.  Emission at most wavelengths, including the infrared, is dominated by a $\sim$120$\arcsec$ radius ``Bright Ring'', which corresponds to $\sim$2 pc at Cas~A's estimated distance of 3.4 kpc \citep{reed95}.  This 30$\arcsec$ thick Bright Ring is formed when ejecta encounter Cas~A's reverse shock and are heated and ionized.  It consists of undiluted ejecta rich in O, Si, S, Ne, Ar, Ca, and Fe \citep{chev78,douv99,hughes00,will03,hwl03,lhw03,morse04}.  Optical, radio, and X-ray observations have revealed the presence of a jet and counter jet in Cas~A.  These jets consist of a bipolar outflow nearly in the plane of the sky \citep{fg96}.  Also visible in the X-ray is a central compact object, presumed to be the neutron star from the progenitor supernova explosion \citep{tan99}.
	
	Cas~A also contains central emission that is not the result of the reverse shock interaction.  This material was first discovered via free-free absorption at low radio frequencies \citep{kas95} and has since been detected in the infrared \citep{rho08, smith09}.  This material was demonstrated to be in substantially different physical conditions than recently shocked material on the Bright Ring through a combination of Doppler analysis and line ratio measurements \citep{smith09}.  The central material is likely photoionized by ultraviolet and X-ray emission from the Bright Ring \citep{hs84, hf98, smith09}, and is relatively poorly studied and understood since it is only visible at select wavelengths.  These ejecta are usually referred to as ``unshocked ejecta'' since they have yet to encounter the remnant's reverse shock.  That is not an entirely accurate label since Cas~A's forward shock, as well as shocks during the supernova explosion itself, have heated this material in the past.  The central emission is ideal for exploring the conditions of the supernova explosion because ejecta interior to the Bright Ring have not yet interacted with the remnant's reverse shock or circumstellar material, leaving them in a relatively pristine state.

	  Recent studies of optical spectra of the explosion near peak light obtained with light echos have led to the observation of weak hydrogen lines, indicating a supernova Type IIb origin for Cas~A \citep{krause08}.  In this scenario, Cas~A's progenitor was the explosion of a red supergiant that had lost most, but not all, of its hydrogen envelope.  The estimated oxygen mass of 1-3M$_{\odot}$ points to a main sequence mass of about 15-25M$_{\odot}$ \citep{young06, vink96}.  X-ray studies indicate a total ejecta mass of less than 4M$_{\odot}$. If one adds to this the mass of the central compact object, presumed to be a neutron star \citep{chak01}, Cas~A's progenitor had a total mass of about 6M$_{\odot}$ immediately prior to the supernova explosion.

Spectropolarimetric observations of supernovae have shown that all observed core collapse supernovae contain intrinsic polarization, indicating that there is a departure from spherical symmetry \citep{whe05}.  An axis-symmetric geometry, perhaps induced by jets, can be used to explain some features in some core collapse supernovae, but significant departures from axial symmetry are needed to explain most observations \citep{wang08}.

\subsection{Previous 3D Studies} Global mappings of Cas~A have been carried out in the optical, infrared, and X-ray.  In the optical, 3D Doppler reconstructions of the ejecta geometry primarily used S and O emission lines \citep{law95,reed95} and showed that ejecta on the Bright Ring lie on a spherical shell but do not uniformly fill that shell; most of the ejecta lie nearly in the plane of the sky.  They also observe that the center of expansion of the ejecta is offset from the geometrical center of the spherical shell by $\sim$0.36pc, indicating that the ejecta are not travelling at the same velocity in all directions with respect to the central compact object.  This is consistent with previous results which indicated a non-spherical expansion for the ejecta \citep[e.g.][]{braun87, will02}. The 3D reconstructions give us a selective ``snapshot'' of ejecta in the sense that only material that has recently encountered the remnant's reverse shock will emit strongly in the infrared and especially in the optical and X-ray.  Emission from material that has yet to reach the remnant's reverse shock is not visible in the X-ray and optical. Currently, ejecta must be travelling at $\sim$5000 km~s$^{-1}$ in order to be encountering the remnant's reverse shock \citep{pf07}.

\cite{delaney09} utilized a spectral mapping of Cas~A from the \emph{Spitzer Space Telescope} and the \emph{Chandra X-ray Observatory} to make a 3D infrared and X-ray map of the remnant\footnote{Movies showing this 3D structure are available at http://chandra.harvard.edu/photo/2009/casa2/animations.html}.  They found a similar distribution of ejecta to that seen in the optical and consistent with a model in which the remnant's reverse shock is a nearly spherical structure $\sim$1.5 pc in radius that is offset from the geometrical center of the remnant both in projection and along the line of sight.  

X-ray Si/S and O emission is observed to be co-located in most regions \citep[e.g.][]{ennis06} in both the X-ray and infrared.  This indicates that the two layers have very similar velocities (less than 80 km~s$^{-1}$ difference) so that they arrive at the remnant's reverse shock at approximately the same time.  However, strong evidence of elemental differentiation is found in some directions in the X-ray \citep[e.g.][]{hughes00}, the optical \citep{fesen06}, and the IR \citep[e.g.][]{smith09,ennis06}, which was likely caused by the different layers of the star being ejected at different velocities in those directions, thus encountering the remnant's reverse shock at different times.  In some regions, only material associated with the Si/S layer is currently encountering the remnant's reverse shock, indicating that that layer has the ``correct'' velocity of $\sim$5000 km~s$^{-1}$.

In other directions, Fe is currently seen at the remnant's reverse shock, indicating that the Si/S and O layers may have a large enough velocity to have already passed through the remnant's reverse shock and ionized up to states which are not detectable in the X-ray.  \cite{ennis06} found regions where only Ne and O are encountering the reverse shock, indicating that the Si/S layer is travelling slowly enough that it has yet to reach the remnant's reverse shock.  Differentiation was also observed by \cite{smith09} in the form of variations in the Ar vs. O + Ne abundances.  The velocity separation between the various layers needs to be several hundreds of km~s$^{-1}$ in order for the layers to reach the reverse shock years apart and produce the differentiation observed. 

We emphasize that we can only observe mixing or separation in \emph{velocity space}.  Simulations predict that the relevant nucleosynthetic layers will be $<$ $10^{11}$ cm thick prior to the explosion \citep[e.g.][]{jog09}. If two such layers are ejected with different velocities, their physical separation will grow with time, and we can detect them individually as they sequentially encounter the reverse shock.  However, if the layers were ejected at the same velocity, they would still form  adjacent $10^{11}$ cm thick shells as they encountered the reverse shock, and we could not separate them.  In addition to the overall shell velocity, there is likely small-scale turbulence which would stretch and broaden clumps and filaments to their observed widths of $\leq$1$\arcsec$ ($\sim$ $10^{16}$ cm).  If this turbulence also mixes the shells, but does not separate them in velocity space, then the layers will encounter the reverse shock at the same time/place.  Thus, if we see separate layers, we know there was a significant velocity difference between them.  If we do not see separate layers, then either their velocities were the same, or they were physically mixed;  we cannot separate those two situations. 

\subsection{Interior Emission}  Previous IR observations also contain information about ejecta that have yet to encounter the remnant's reverse shock \citep[e.g.][]{smith09}.  These are visible because some IR ionic lines, like the 34.81$\mu$m [Si~II] line, will be photoionized by X-Rays and UV light from the Bright Ring even if they have yet to reach the remnant's reverse shock.  They can therefore be visible interior to the remnant's reverse shock or far beyond it if the ejecta passed through it decades ago \citep{chev03}.
	
In the case of Cas~A, these interior ejecta give it a filled center appearance \citep{smith09}, as opposed to being dominated by the Bright Ring. \cite{delaney09} show that this interior emission is organized into a ``Thick Disk'' structure, tilted at $\sim$70 degrees from the line of sight.  The material is moving perpendicular to that plane at $\sim$2500 km~s$^{-1}$, indicating that it is only about half-way to the remnant's reverse shock. Figure 1 illustrates the relationship between the remnant's reverse shock, the Bright Ring, and the interior ejecta.

We present an analysis of a new, higher resolution Spitzer mapping of the ejecta towards the center of Cas~A.  In $\S$2 we present the Spitzer observations.  In $\S$3 we discuss the methods used in our analysis and we describe those results in $\S$4.   $\S$5 contains a discussion of the physical implications.

 \section{Spitzer Observations}The Spitzer Infrared Spectrograph (IRS) was used on August 30, 2007 to spectrally map select relatively bright regions of Cas~A.  This paper will address only the central map whose location is shown in Figure 2; a followup paper will address the other regions.  High-resolution spectra (R$\sim$600 for all wavelengths) were taken between 10-20$\mu$m and 20-35$\mu$m using the Short High (SH) and Long High (LH) modules respectively.  The full-width-half-max of the lines at this spectral resolution is about 0.06$\mu$m at 35$\mu$m and about 0.02$\mu$m at 13$\mu$m.  This represents an improvement in spectral resolution of a factor of $\sim$6 over the earlier observations of \cite{delaney09}.  The LH data were taken in a single large map with 3x15 pointings using a 61 second exposure at each position.  The SH data were taken with 6x15 pointings using a 31 second exposure at each position.  The mapped area ranged from 54$\arcsec$x40$\arcsec$ (LH) to 48$\arcsec$x36$\arcsec$ (SH) at a spatial resolution of $\sim$1.25$\arcsec$ and $\sim$2.5$\arcsec$ respectively.

	The spectra were reconstructed at each slit position, the background was subtracted, and 3D cubes were created using the S17 version of the IRS pipeline and the CUBISM software package \citep{smith07}.  The statistical uncertainties for each line of sight were calculated using standard error propagation of the BCD level errors from the standard IRS pipeline.

In general, our uncertainties were limited by the undersampling of the IRS modules, which is worst at the short-wavelength end.  This is a systematic error that exists in the wavelength calibration data themselves.  It limits our obtainable absolute wavelength accuracy to roughly 1/2 of a spectral bin, or about 100 km~s$^{-1}$, although the relative wavelengths for a given line can be measured with higher accuracy.

\section{Data Analysis}

Cas~A's infrared spectrum is dominated by bright ionic emission lines as shown in Figure 3.  The LH observation contains lines from [O~IV] and/or [Fe~II] at 25.9$\mu$m, [S~III] at 33.48$\mu$m, and [Si~II] at 34.81$\mu$m.  The SH observation has a [S~IV] line at 10.5$\mu$m, [Ne~II] at 12.8$\mu$m, and another [S~III] line at 18.7$\mu$m.  The lines observed in the LH module typically have peak fluxes from 200-4000 MJy sr$^{-1}$, with an rms noise of $\sim$15.  Lines in the SH module have peak fluxes from 12-250 MJy sr$^{-1}$ and a typical rms noise of $\sim$8.  We also tentatively identify the 17.94$\mu$m [Fe~II] line with 2$\sigma$ significance when we spatially bin all pixels over the entire central region.

Observed emission near 25.9$\mu$m could be from either the 25.89$\mu$m [O~IV] line or the 25.98$\mu$m [Fe~II] line.  In order to differentiate betwen the two lines, we compared the Doppler structure of the 25.9$\mu$m line to that of the [Si~II] and 33.48$\mu$m [S~III] lines for several lines of sight.  In Figure 4 we display the results for one line of sight, showing the velocity structure of the [Si~II] line at 34.81$\mu$m, the [S~III] line at 33.48$\mu$m, as well as the 25.9$\mu$m line under the assumption that it is either composed entirely of either [O~IV] or [Fe~II] emission.  We obtain an excellent match under the assumption of [O~IV], but a poor match under the assumption of [Fe~II].  The mismatch in Doppler structure under the assumption of [Fe~II] cannot be due to [Fe~II] simply having different velocities than [Si~II] and [S~III] since the [Fe~II] ejecta would have to be moving more rapidly than [Si~II]/[S~III] on the front side of the remnant and more slowly on the back side in order to produce the observed spectrum.  Thus, it is clear that the velocity structure is consistent with the line being composed entirely of [O~IV].  We find no lines of sight that are consistent with having a substantial contribution from Fe.  We assume for the remainder of this paper that the 25.9$\mu$m line is entirely due to [O~IV] emission.

Th above analysis is based on the assumption of 25.8913$\mu$m for the rest wavelength of the [O~IV] line \citep{feucht97}.  This differs from the earlier value of 25.913$\mu$m \citep{ff83}.  \cite{feucht97} note that their results substantially improve upon the accuracy of previous values which were primarily based on energy level differences reconstructed from UV and optical spectroscopy.

Although [Si~II] and [O~IV] match up well along single lines of sight, the relative strength of [Si~II] and [O~IV] varies considerably from place to place.  Therefore, the total line shapes from all interior emission are considerably different for the two ions as shown in Figure 5.

\subsection{Doppler Deconvolution}  After background subtraction, we performed a Doppler deconvolution of the spectral lines from each ion separately using a spectral CLEAN algorithm \citep{ding99} for each line of sight.  SH data cubes were binned 2 by 2 pixels (approximately 2.5$\arcsec$ by 2.5$\arcsec$) to increase the signal to noise ratio.  A careful deconvolution is preferable to simpler techniques like measuring the peak wavelength of emission lines because the spectral CLEAN algorithm is able to separate partially blended components.  An example of a CLEANed spectrum is shown in Figure 6.  Note that flux in neighboring spectral bins was assumed to be from the same Doppler component.  In this case we combined the bins and determined the peak wavelength of the Doppler component by taking a weighted average of the wavelengths of the combined bins.

  The spectral CLEAN was applied to each spatial pixel that had a signal greater than 3 times the off-line rms.  We note that at our spatial resolution ($\sim$2.5$\arcsec$), we may be binning over many individual knots.  At the remnant's reverse shock ejecta knots which have typical sizes as small as 0.2$\arcsec$-0.4$\arcsec$ are observed \citep{fesen01}.  We do not know the spatial size of any clumping in the interior ejecta.

	Uncertainties in Doppler velocity for a given Doppler component were determined by applying the spectral CLEAN to synthetic line data with a realistic range of signal to noise ratios, and using actual line free data for the noise model.  From these simulations, we determined the rms error in velocity as a function of line strength and location of the line peak within a spectral bin.  In all cases, the uncertainty in velocity for a single, isolated Doppler component was determined to be less than 25 km~s$^{-1}$.  This means that our uncertainties in the absolute velocities are limited by the systematic errors in the calibration of about 100 km~s$^{-1}$ rather than random uncertainties, while the relative velocities for any given line are less than 25 km~s$^{-1}$.  We could not detect two separate components that were within 65 km~s$^{-1}$ of one another in synthetic data. 

We assume that the ejecta have been freely expanding at a constant velocity in order to determine their spatial coordinate perpendicular to the plane of the sky.  This assumption is still valid despite the fact that the ejecta were likely decelerated by shocks during the supernova explosion itself - any deceleration happened at the time of the explosion (that is, near t=0, z=0 where z is the spatial coordinate perpendicular to the plane of the sky) so the behavior is virtually identical to free expansion at a reduced velocity.  We transformed our Doppler velocities to a z axis spatial coordinate in Figure 7, but leave the z axis in velocity units in Figures 8-13. The flux from each component is displayed by varying the transparency; the brightest voxel for a given ionic line is 80\% opaque, while the opacity of all other voxels is linearly scaled downwards as a function of the intensity of the Doppler component.

We note that our results are consistent with the lower spectral resolution results of \cite{delaney09}.  When plotted on the same axes, the ejecta detected in both observations trace out similar structures as shown in Figure 7, although our superior spectral resolution (R$\sim$600 vs R$\sim$100) allows us to detect many details that were previously not observed.

\section{Results} 

\subsection{3D Map}

We plot the Doppler components from the 3 strongest lines in Figures 8-10 - the 25.89$\mu$m [O~IV] line, the 34.81$\mu$m [Si~II] line, and the 33.48$\mu$m [S~III] line.  The other lines are either very weak (in the case of Ne and Fe) or trace out identical structures as other lines from the same element (in the case of the other S lines).  The velocity axis has been stretched by a factor of 1.8 in order to highlight features in velocity space. Due to the low density of the interior ejecta \citep[$\le$100 e$^{-}$ cm$^{-3}$,][]{smith09} we expect line of sight absorption to be minimal.  However, we note that we are likely only observing the very densest ejecta since a small decrease in density will result in a substantial drop in emissivity.  Therefore, it is likely that there is a large amount of undetected interior ejecta present that is at too low a surface brightness to be detected.

The most striking aspect of this emission is that the center of the remnant is not uniformly filled, but consists of distinct structures.  We label the material on the back side of the remnant as the ``Sheet'', the material on the front side of the remnant as the ``Filament Band'', and the material between the two as the ``Bridge''.  The Si and S lines trace out essentially identical structures as seen in Figure 11.

The Filament Band and the Sheet are orientated at $10.4^{\circ}$ and $16.4^{\circ}$, respectively, with respect to the plane of the sky.  This is consistent with the range of $\sim$ $25^{\circ}$ $\pm$ $15^{\circ}$ in orientation across the Thick Disk observed in the low resolution study \citep{delaney09}.

\subsection{Ejecta Structure Asymmetry}

There is a striking front-back asymmetry in the  geometrical structure of the ejecta.  The Si, S, and O ejecta in the Sheet have a very narrow velocity profile - along any given line of sight all of the material has only one Doppler component.  The structure itself is remarkably well formed in that in almost all places it is $<$65 km~s$^{-1}$ thick - the minimum thickness allowed by our observations.  The ratio between the O and Si lines varies considerably - some regions appear almost entirely in one line or the other.  As will be discussed in $\S$5.5, we do not know if this is due to actual elemental abundance variations or variations in line strength due to density and temperature variations.  The structure is nearly continuous except for a hole in the structure in both Si and O (indicated in Fig. 8). 

In the Filament Band the material forms an interwoven filamentary structure.  About half of the lines of sight contain more than one Doppler component.  In general, the filaments are nearly as narrow as possible given our spectral and spatial resolution ($\sim$0.03pc thick) and can be up to $\sim$0.3pc long.  The emission in the Filament Band has a total width in velocity space of roughly 1500 km~s$^{-1}$ along each line of sight, compared to a width of $<$65 km~s$^{-1}$ for the Sheet.  Without high resolution data throughout the interior of Cas~A, we cannot tell whether these structural asymmetries apply to the entire Thick Disk described by \cite{delaney09}.

\subsection{Ejecta Velocity Asymmetry}

There is a substantial difference in the overall velocities of the Sheet and Filament Band regions, on top of the large variations in both intensity and velocity in the various lines.  The intensity weighted average velocity of ejecta in the Sheet and Filament Band are 2900 km~s$^{-1}$ and -2000 km~s$^{-1}$ respectively, as shown by the total line shapes in Figure 5.

The velocities of ejecta in the Sheet vary from $\sim$2000-4600 km~s$^{-1}$, going from East to West.  The strongest concentration of Si emission is at $\sim$3000 km~s$^{-1}$, while the O is spread more evenly over the velocity range.

The average velocity of the ejecta in the Filament Band region ranges from -1500 to -3800 km~s$^{-1}$, going from West to East.  The strongest concentration of Si emission in the Filament Band is at $\sim$-1600 km~s$^{-1}$, much slower than the Si in the Sheet.  The O has a larger spread in velocities, but there is substantial O flux at velocities around -1500 km~s$^{-1}$, 500 km~s$^{-1}$ slower than any O emission in the Sheet.

Put together, our observations indicate that ejecta in the Filament Band region are typically travelling $\sim$900 km~s$^{-1}$ more slowly than ejecta in the Sheet region.

\subsection{Radial Velocity Profile Asymmetry}

We define the radial velocity profile across nucleosynthesis layers as the mass of the ejecta that are traveling at a given radial velocity for each element.  Examples of two radial velocity profiles are shown in Figure 13.  We can characterize the radial velocity profiles of the original supernova explosion by determining if different nucleosynthetic layers are separated in velocity space.  If we observe that Si and O are separated, then we know that Si and O were ejected at different velocities and thus have different radial velocity profiles.

We can qualitatively see a variety of radial velocity profiles in our data.  The O and Si in the Sheet appears to be strongly overlapping in Figure 12 where [O~IV] and [Si~II] emission is plotted on the same axes, indicating that two elements were ejected at the same velocity.  We quantify the velocity separation between Si and O in the Sheet as follows: for each Si component with flux greater than 100 MJy sr$^{-1}$, we found the nearest O component along the same line of sight provided that its velocity was within 1000 km~s$^{-1}$ of the Si velocity and its flux was greater than 100 MJy sr$^{-1}$.  We plotted the Si velocity vs O velocity from all lines of sight in Fig. 14.  If the elemental layers had identical velocity profiles, the velocities would be equal.  We find that the slope of the best fit line of the combined data and forced to pass through the origin is 1.015 $\pm$0.0025.  This corresponds to the O having a mean velocity 45 $\pm$14 km~s$^{-1}$ greater than the Si at the average position of the Sheet.  Since any separation between Si and O is less than the $\sim$100 km~s$^{-1}$ uncertainty induced by systematic errors, we conclude that the mean Si and O velocities are identical, within that limit, when both elements are visible.

The rms scatter of the points about the best fit line is 75 km~s$^{-1}$.  This is much larger than the random uncertainty in velocity, which is always less than 25 km~s$^{-1}$ in both O and Si velocities.  This indicates that the scatter is not statistical or instrumental in nature, but is a real variation in the supernova ejecta itself.  However, this scatter is very small - it represents a variation of only $\sim$1\% in the average ejecta velocity in the Sheet.  

Turning now to the Filament Band, in roughly half of the filaments we find no separation between the O and Si velocities.  The other half of the filaments are composed almost entirely of either O or Si.  The filaments have characteristic separations of order 500 km~s$^{-1}$.  This separation cannot be due to contributions from both [O~IV] and [Fe~II] since we would be able to individually resolve both the [Fe~II] and [O~IV] lines if both elements were present.  We are also not mistaking the [O~IV] emission for [Fe~II] since the difference in rest wavelengths would result in a velocity change of $\sim$1000 km~s$^{-1}$.  Thus, the filaments would still be separated even if we were detecting Fe emission.

We do not attempt to directly compare the Si and O velocities in the Bridge or Filament Band overall, since the ejecta from each element in those regions are often in completely different structures.

We note that the radial velocity profile for every line of sight in the Sheet must be very strongly peaked in both Si and O because we observe one and only one velocity for both elements.  However, the radial velocity profile in the Filament Band is much broader since we can see a range of velocities in many lines of sight.

\subsection{Line Fluxes}

We determine line ratios of the 18 and 33 $\mu$m [S~III] lines for several lines of sight.  These lines can be used to determine the density of the ejecta (assuming that they are at a high enough density and temperature) by balancing the collisional excitation and de-excitation rates as well as radiative transitions into and out of the relevent energy levels \citep[e.g.][]{ost06}.  \cite{smith09} used this diagnostic on data that had not been deprojected and found that all lines of sight in our field of view had ejecta with electron densities $<$100 cm$^{-3}$, the lower limit of this density diagnostic.  We attempted to identify any Doppler components with densities $>$100 cm$^{-3}$ by determining the [S~III] ratio for the 5 Doppler components with the strongest 18$\mu$m line.  However, in all cases we found that the electron density is $<$100 cm$^{-3}$, confirming the results of \cite{smith09}.

We give the integrated line fluxes for two Doppler components from the same line of sight in Table 1 as an example of typical values.  Note the variation in the [S~III] line ratio between the two components, which demonstrates that it is often necessary to deconvolve the data before attempting to extract information from line ratios.  We find that this [S~III] line ratio varies between roughly 0.02 and 0.4 for deconvolved components, with most ratios around a value of 0.05.  Since we do not yet have the appropriate models to determine physical conditions from these line ratios, we defer further discussion of line ratios to a future paper.  We address the need for additional modeling in $\S$5.5.

\subsection{[Ne II] Map}

Although the 12.8$\mu$m [Ne~II] line is too weak to extract a substantial number of individual Doppler components, we can map the [Ne~II] flux distribution over our field of view.  We compare this map to a 25.9$\mu$m [O~IV] flux map in Figure 15.  Naively, we expect substantial similarities between the two maps since the elements come from the same nucleosynthetic layer.  However, the two maps do not show a strong correlation.  We briefly address potential reasons for these differences in $\S$5.5.

\section{Discussion}

\subsection{Supernova Model Background} One of the great outstanding problems in theoretical astrophysics is the basic nature of core-collapse supernova explosions.  In contrast, the structure of the star before the supernova explosion is relatively similar for all models.  As the massive star fuses different elements during hydrostatic burning, it should produce denser and denser concentric nucleosynthesis layers, forming the classic ``onion-skin'' model of the star.  We concentrate on the central layers of the star - the dense Fe/Ni core, the Si/S layer immediately above the core, and the O/Ne layer above the Si/S.  

Any mixing between the layers during the supernova explosion itself could be caused by either partial explosive O burning \citep{chev79} or mixing between layers caused by large-scale Rayleigh-Taylor instability fingers created by shocks during the supernova explosion \citep{wink91}.  However, our observations indicate that the Rayleigh-Taylor filament scenario is more likely since we observe filamentary structures in ejecta which have not yet encountered the remnant's reverse shock and there is no obvious way that partial explosive O burning could create Si filaments radially flanked by O filaments.  Therefore, we do not further discuss any scenarios based on mixing by explosive nucleosynthesis.

Although the initial conditions are well understood, the nature of the piston responsible for the explosion itself is not, with most groups proposing neutrino-driven shocks, while others utilize diffusive, magnetic buoyancy or neutrino-bubble instabilities \citep{janka07}.  Regardless of the exact nature of the piston, many predictions about the early shock structure of the supernova explosion can be made.  As the primary piston drives outwards, it causes a forward shock and eventually sweeps up enough material in the star to form a reverse shock \emph{within the star itself} \citep{hw94}.  This ``Explosion Reverse Shock'' forms about $10^{3}$-$10^{4}$ seconds after the beginning of the supernova explosion and takes $\sim$ $10^{2}$ seconds to reach the center of the star \citep{jog09}.  This is different than the ``Remnant Reverse Shock'', which formed
$\sim 10^2$ years later in Cas~A when the forward shock swept up enough material to cause the Remnant Reverse Shock to separate \citep{miles09}, and has not yet reached the center of the remnant.  Figure 16 illustrates the distinction between these two reverse shocks.

The Explosion Reverse Shock forms in the outer layers of the star and propagates toward the center of the star, forcing the less dense outer nucleosynthetic layers into the denser layers deeper within the star.  This can cause mixing between the layers and potentially forms filaments from Rayleigh-Taylor instabilities \citep{hw94}.  The amount of mixing and degree of filamentation depends on the the speed of the reverse shock, which can vary by roughly an order of magnitude in models of stars with different masses \citep{jog09}.  A strong and fast Explosion Reverse Shock can cause complete mixing between many layers and prevents the production of filaments because it sweeps by so quickly that filaments do not have time to grow.  The signature of this phenomenon is well mixed sheets of ejecta.  On the other hand, a slower Explosion Reverse Shock can result in large filaments because it moves slowly enough that the filaments have time to develop.  A very weak shock would not be strong enough to cause much mixing at all between most elements, leaving the nucleosynthetic layers spatially separated \citep{jog09}.

The Explosion Reverse Shock simultaneously modifies the radial velocity profile of the ejecta across concentric nucleosynthetic layers. Elements that have been well mixed by the Explosion Reverse Shock (indicating a strong and fast Explosion Reverse Shock) should have nearly identical velocities upon ejection, while unmixed layers (indicative of a weaker and slower Explosion Reverse Shock) can have velocities that differ by 1000s of km~s$^{-1}$. This is a key distinguishing feature between supernova models.  Some models predict that the Si/S and O layers will have essentially identical velocities, while other models predict that the layers can be separated by 1000s of km~s$^{-1}$ \citep[Fig. 14 - ][]{jog09,kifonidis06}.  Even within very similar models, the separation between layers can be a function of the initial conditions within the explosion - \cite{jog09} predict that the Si/S and O layers will have nearly identical velocities for 25M$_{\odot}$ solar metallicity stars, while they will be separated by $\sim$1000 km~s$^{-1}$ for 15M$_{\odot}$ solar metallicity stars. 

An alternative to spherical shocks is a jet-induced supernova explosion \citep[e.g.][]{burrows07}.  In this scenario, the explosion is dominated by MHD jets created in rapidly rotating stars.  When these stars explode, the jets induce a bipolar outflow and create powerful bow-shocks as they move through the star.  These transverse shocks eventually collide at or near the equator of the star, leading to a torus of ejecta about the star's equator \citep[e.g.][]{kho99}.  Stars with moderate rotation may have supernovae with both spherical and weak transverse jet induced shocks \citep{burrows07}.

\subsection{Nature of the Explosion}

We can use supernova explosion models as a guide to which physical properties may influence the observed asymmetries.  The fundamental cause of the ejecta structure, the ejecta velocity, and the radial velocity profile asymmetries described in this paper may be variations in the Explosion Reverse Shock, which were potentially caused by the variations in the forward shock.  In this model, the Explosion Reverse Shock was very strong and moving very quickly in some directions, leading to the Sheet structure where the elements are mixed in velocity space and no filaments are seen.  In other directions it was slower, leading to filaments composed of both Si and O.  In yet other directions it was very weak and slow, leading to well separated filaments.

Another potential source of asymmetries are those that form in the forward shock within the first $\sim$100 milliseconds in the models of \cite{burrows07} as well as standing acoustic shock instability (SASI) models of \cite{blondin03}.  These instabilities allow the initially steady-state, spherically-symmetric forward shock to become highly asymmetric in just a few crossing times \citep{blondin03}.  The origin of these instabilities is the response of the post shock pressure to changes in the forward shock radius and happens while the forward shock is roughly stationary and still very near the core of the star.  If the pressure in one regions becomes only slightly higher than equilibrium, it will push the forward shock outward.  The preshock ram pressure drops with increasing radius, so the outward shock displacement leads to smaller pressure behind the forward shock.  But, if the postshock pressure radial profile is steeper than the preshock ram pressure profile, positive feedback and a standing acoustic wave are produced \citep{blondin03}.  This leads to a forward shock with low-order asymmetry.  Presumably, the Explosion Reverse Shock would be strongly affected by this asymmetry when it separates from the forward shock.

Jet-induced supernova explosions do not appear to be an attractive alternative for explaining our observations of Cas~A. The distinct, tilted front/back structures that we report here and are part of the \cite{delaney09} ``Thick Disk'', are not orientated correctly to be formed as a torus in a jet induced supernova explosion.  They are nearly in the same plane as the jet/counter-jet, not perpendicular to it as the jet models require.  While the jets do produce a slight bipolar asymmetry in the ejecta, there is no obvious way in which the jets cause most of the ejecta asymmetries described in $\S$4.  Furthermore, Cas~A's jets do not appear to have enough kinetic energy in order to cause the supernova explosion \citep{laming06}.

Wheeler et al. (2008) propose that the structures normally called the ``jet'' and ``counter-jet'' are not the main jets, but rather secondary instabilities.  In this model, the jets responsible for the explosion are two notable Fe blowouts located in the Southeast and Northwest of the remnant.  However, the 3D reconstructions of \cite{delaney09} show that these two blowouts do not form an axis.  The Fe blowouts also are not nearly perpendicular to the ``Thick Disk'' as expected in a jet induced explosion.

We do not further consider jets as the source of Cas~A's asymmetries any further since both jet-induced scenarios do not seem to be plausible.

Other models can also produce ejecta asymmetries from rotation without using jets to induce the supernova explosion.  However, the published results of such models \citep[e.g.][]{kifonidis06} do not document any of the key asymmetries in ejecta structure, velocity, and radial velocity profile that we need to compare with the current observations.  It is clear that one key to understanding and reproducing the asymmetries is to have models that predict the average ejecta velocity as well as the radial velocity profile \emph{as a function of direction}.

\subsection{Relationship Between the O and Si/S Layers}

X-ray observations have led to the suggestion that Cas~A's nucleosynthetic layers have undergone large-scale overturning in some regions, causing less dense layers to be interior to layers which originated closer to the star's core \citep[e.g.][]{hughes00}.  In most directions we find no evidence of this overturning as the the O and Si/S layers are nearly perfectly correlated in velocity space (see Figs. 12 and 14).  This is consistent with IR and X-ray observations that indicate that Si/S and O emission is co-located on much of the Bright Ring \citep[e.g.][]{ennis06}.   However, in part of the Filament Band we do find substantial separation between nucleosynthetic layers.  This is roughly consistent with the separation between layers seen in the X-ray, but does not correspond to a simple overturning of the layers since the O-layer is observed on \emph{both} sides of the initially denser Si/S layer in velocity space.

While this intertwining of O and Si/S layers may be evidence of some sort of mixing between nucleosynthetic layers in some parts of the star, it does not apply in other directions.  Future supernova explosion models that better address the asymmetries seen in Cas~A may shed light on this interesting phenomenon.

We also note that our results are similar to those of \cite{fesen06}.  Based on Hubble Space Telescope observations of select regions of the remnant, they also concluded that there was substantial spatial variation in the degree of mixing of the layers in Cas~A.  Their data consisted of knots composed of lighter elements that originated in the outer layers of the star; our results show that the variability in mixing remains even down to the denser interior layers.

\subsection{Velocity Offset} 

Previous authors observing ejecta in the optical found that the center of expansion of the ejecta was offset along the line of sight from the geometrical center of the partial spherical shell (caused by ejecta interacting with the Remnant Reverse Shock) by $\sim$770 km~s$^{-1}$ \citep{reed95}.  Our new IR results are roughly consistent with this result - we find an offset of $\sim$900 km~s$^{-1}$ along our line of sight in the same direction.  However, Reed et al. (1995) speculated that this was due to a difference in density of the circumstellar material between the back and front of the remnant.  This is inconsistent with our data - the interior ejecta visible in the infrared are unaffected by the circumstellar material because they have not yet encountered the Remnant Reverse Shock.  Thus, we believe that this velocity offset is the result of an asymmetry in the supernova explosion itself rather than an asymmetry in the circumstellar material.

\subsection{Interior Conditions} 

Our observations raise an interesting puzzle with respect to Fe.  We do not definitively detect any Fe in the interior, despite Fe II, Fe III, and Fe VII lines within the wavelengths accessible to \emph{Spitzer}'s IRS module.  This could either be due to a lack of Fe in the interior of the remnant or because the Fe present is not in the correct physical conditions to emit detectable lines.  We believe that the latter scenario is more likely since recently shocked Fe is observed on the Bright Ring \citep[e.g.][]{hughes00,ericksen09} and we know of no mechanism which would force all the Fe, and only the Fe, to be ejected only in a narrow torus.

One possible solution to this puzzle is that the Fe is at lower density than the observed Si and O.  There are multiple explanations for how this may occur, but we will only discuss one here.  Although the Ni/Fe layer is initially more dense than the Si/S and O layers, it may not have experienced the same modifications to its density distribution as the outer layers.  For example, if the Explosion Reverse Shock lost most of its energy before it reached the Fe/Ni core, it would not cause the Rayleigh-Taylor filamentation that is likely responsible for the dense knots of Si/S/O that we observe in the interior.  The phenomenon of density enhancements to outer layers but not the core is seen in some of the models of \cite{jog09}.  Fe may not be observable without this density enhancement.

A similar puzzle arises with respect to the Si and O lines.  In some regions with coherent structure like the Sheet, we see emission which is almost entirely O or Si.  These variations in line strengths could be due to either variations in local abundance ratios between the elements or density and temperature variations.  We find that the brightness of a region in Si is uncorrelated with how bright it is in O and vice versa.

Similarly, we observe that the [Ne~II] and [O~IV] maps show substantial differences although both elements came from the same nucleosynthetic layer.  We speculate that this is because the emissivity of the two lines is a function of density and temperature, and the variation in line flux is therefore reflecting a variation in physical conditions.

We look forward to future models which balance photoionization rates (as opposed to the collisional ionization rates used in the usual 18/33 $\mu$m [S~III] diagnostic) in order to determine line ratios as a function of temperature and density.  These models should be able to probe the low temperatures and densities present in Cas~A's center.  Not only will we be able to better address the puzzles presented above, but we will also be able to better constrain the temperature and density of the interior ejecta. 

\subsection{Predictions for the Next 30-500 Years}

Cas~A's appearance in the X-ray is dominated by ejecta which have recently encountered the Remnant Reverse Shock.  Thus, the central ejecta arriving at the Remnant Reverse Shock will mark a transition after which Cas~A will contain bright central emission in the optical and X-ray.  This situation will be analogous to the supernova remnants N132D \citep{blair00} and Puppis A \citep{wink85}.  Like in N132D, we will observe a ring of ejecta with arcs and clumps of bright, recently shocked ejecta interior to the ring.  We note that the two scenarios are not an exact analogue since N132D's appearance is dominated by its forward shock interacting with the surrounding environment.  The remnant will still be a shell morphology remnant (since the ring will be limb brightened), but a substantial fraction of the overall X-ray emission will be from shocked ejecta in the interior.  However, like in Puppis A we expect the central ejecta to be O rich since our IR observations detect strong O lines in the interior ejecta.

We can make explicit predictions about when the central ejecta should encounter the Remnant Reverse Shock with our knowledge of the current velocity structure of the ejecta and by assuming, in the limiting case, that the Remnant Reverse Shock is roughly stationary in the current epoch \citep{morse04}.  On the back side of the remnant, we expect the Sheet to begin arriving at the Remnant Reverse Shock in slightly under 30 years at about RA: 23:23:25 and Dec: 58:48:5.  The X-ray and optical emission may initially be dominated by O group emission, since the part of the Sheet with the highest velocity is dominated by O emission in our observations.  The greatest concentration of Si group emission on the back side of the remnant should arrive at the Remnant Reverse Shock in approximately 220 years. If the Remnant Reverse Shock actually begins moving inward in the external reference frame during this period, then these times will be shorter.

On the front of the remnant, the ejecta in the Filament Band will begin to arrive at the Remnant Reverse Shock in about 100 years.  The ejecta in the Filament Band with the largest velocity are the overlapping filaments, so the emission will initially be strong in both O and Si.  The separated filaments will begin to arrive at the reverse shock in about 260 years.  The emission will intially be dominated by O.  The Si group emission should begin about 500 years from the present time in this direction and will be followed by more O group dominated emission several decades later.  We note that the substantial separation in arrival times of the O and Si group in this direction is consistant with the \cite{ennis06} result in which emission in some regions are observed to contain almost exclusively O and Ne, with little Si.

\section{Conclusions}

We create a 3D reconstruction of the central ejecta of Cas~A at unprecedented spectral resolution using photoionized infrared ionic lines.  We observe a large number of asymmetries that are most likely caused by asymmetries in the supernova explosion itself rather than the circumstellar environment.  Si and O emission with nearly identical velocities are seen in co-located sheets less than 100 km~s$^{-1}$ thick on the back side of the remnant.  Toward the front, by contrast, we observe filaments with both Si and O present, while along different lines of sight we observed well-separated Si and O filaments that are roughly consistent with X-ray observations.  The average velocity of all ejecta varies strongly as a function of direction.   We observe that the interior emission is offset by $\sim$900 km~s$^{-1}$ along our line of sight as was previously observed in the optical.  However, we do not believe that this asymmetry was caused by the circumstellar environment because the interior ejecta can not be affected by the ISM until they reach the forward shock.  We hypothesize that the asymmetries could be produced by asymmetries in the Explosion Reverse Shock.

Photoionization models are required in order to determine the density, temperature, and ionization state of the center of the remnant.  These models are likely to be produced in the near future and will allow us to further address the central conditions of the remnant.  One key question to be answered is whether or not the lack of a detection of any Fe lines indicates a lack of interior Fe, or that the Fe is present, but not in the correct physical state to produce observable lines.

Finally, we note that Cas~A will provide an even more fertile ground for future observations as the interior ejecta encounter the front and back Remnant Reverse Shocks, starting in $\sim$30 years.

\acknowledgements

This work is based on observations made with the Spitzer Space Telescope, which is operated by the Jet Propulsion Laboratory, California Institute of Technology under NASA contract 1407.  This work was supported in part by NASA/SAO Award No. AR5-6008X and NASA/JPL through award 1265552 to the University of Minnesota.

K.~I. would like to thank Alex Heger for valuable conversations concerning supernova explosion physics.  K.~I. would also like to thank Kris Eriksen for his insights concerning photoionization models of the interior emission in Cas~A. We also appreciate the useful comments of the referee.

\newpage
\begin{deluxetable}{cccccc}
\tablecaption{Integrated line flux from front and back emission for a 2.5$\arcsec$ x 2.5$\arcsec$ example region (23:23:31, 58:48:43). Typical uncertainties in line flux are $\le$15\%.}
\tablehead{ & \colhead{[S IV]} & \colhead{[S III]} &
\colhead{[O IV]} & \colhead{[S III]} & \colhead{[Si II]} \\
\colhead{ } & \colhead{10.5 $\mu$m} & \colhead{18.7 $\mu$m} & \colhead{25.9 $\mu$m} & \colhead{33.5 $\mu$m}& \colhead{34.8 $\mu$m} }
\startdata

Back Integrated Flux & 0.0114 & 0.0960 & 1.25 & 0.632 & 4.93 \\
($10^{-17}$ W m$^{-2}$) & & & & \\
& & & & & \\
Front Integrated Flux & 0.0298 & 0.391 & 2.12 & 1.02 & 6.37 \\
($10^{-17}$ W m$^{-2}$) & & & & \\

\enddata
\end{deluxetable}
\clearpage

	
\begin{figure*}
    \centering
    \includegraphics[width=0.5\textwidth]{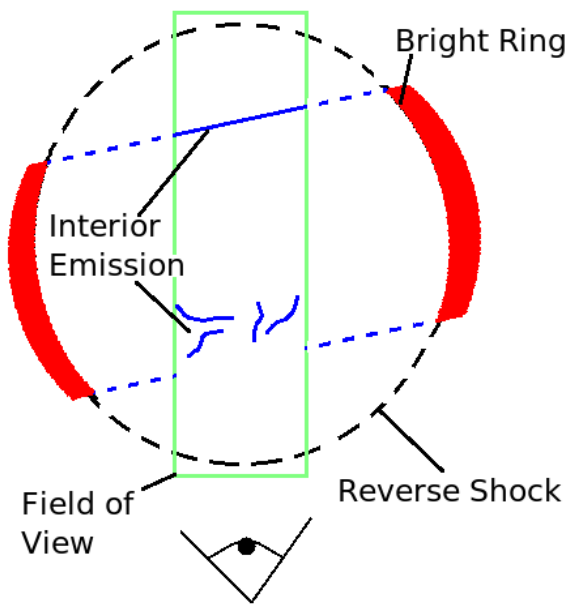}

\caption{\small 2D projection cartoon of the shock and ejecta structure showing the connection between the observations presented here and the model of \cite{delaney09} as viewed from the top.  The remnant's reverse shock is a nearly spherical structure, while the ejecta are flattened nearly perpendicular to the plane of the sky.  Only part of the reverse shock is observable, while the sections of the remnant's reverse shock that are not currently encountering ejecta are not currently observable (dashed black).  Ejecta that are currently encountering the remnant's reverse shock will be visible as mixed X-ray, IR, and optical emission (red), while ejecta interior to the remnant's reverse shock will only be visible in select IR lines (blue).  The approximate field of view of the current observations is indicated (green) within which are the different structures as discussed in $\S$4.
\normalsize }
    \label{fig:large1}
\end{figure*}


\begin{figure*}
    \centering
    \includegraphics[width=0.4\textwidth]{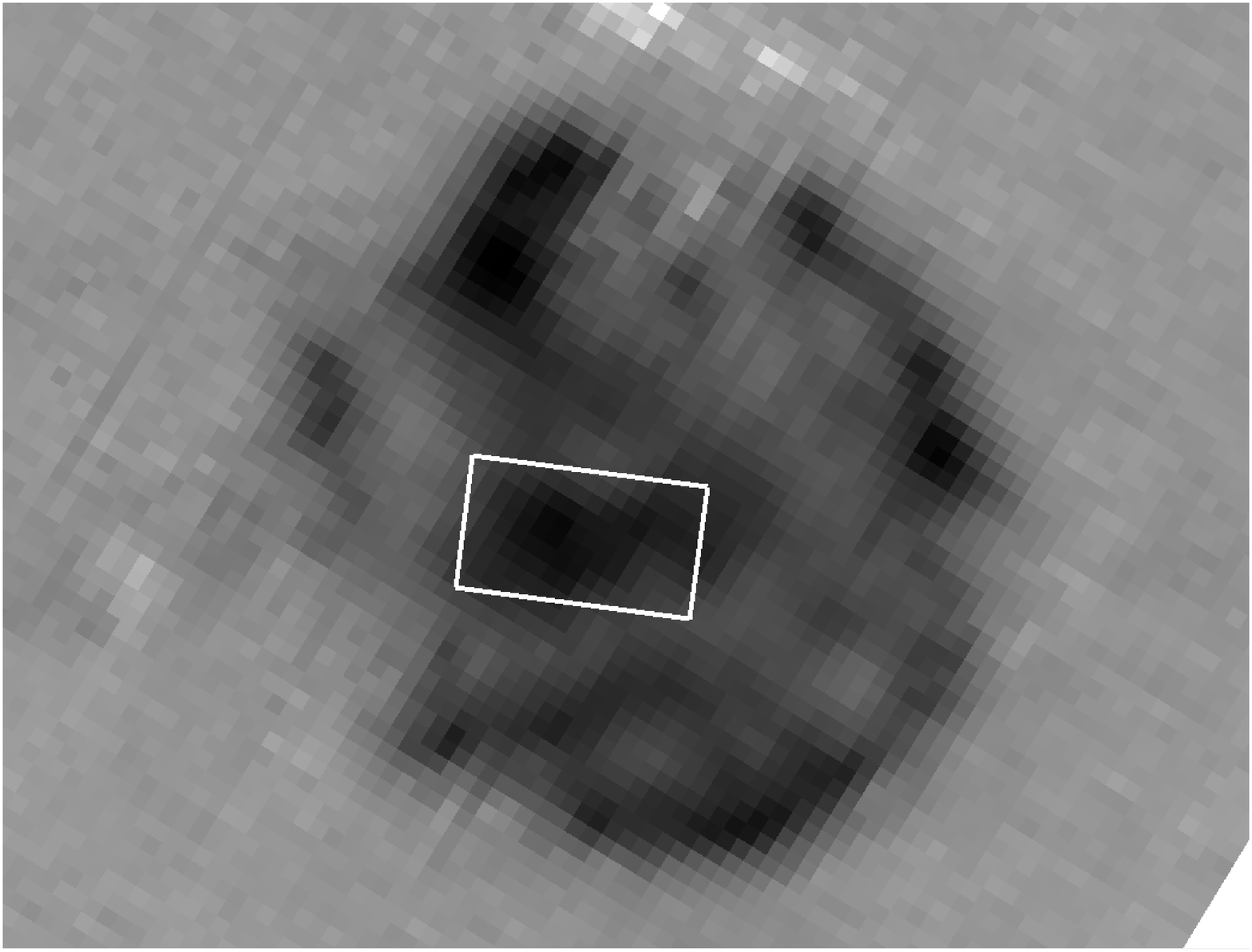}
    \includegraphics[width=0.4\textwidth]{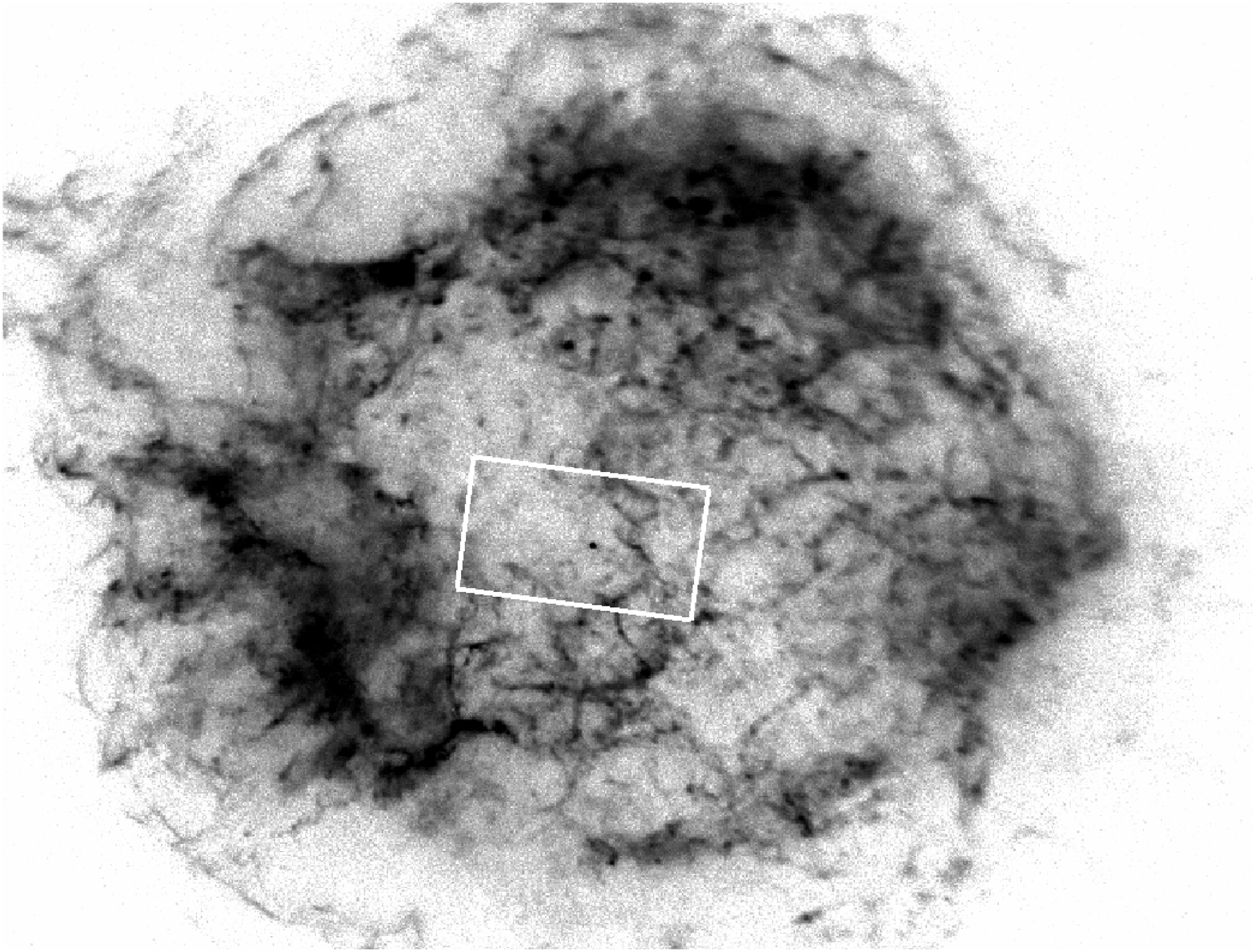}

\caption{\small 34.81$\mu$m [Si~II] \emph{Spitzer} IRS map and X-ray Si \emph{Chandra} map of Cas~A.  The region of high resolution data discussed in this text is indicated.
\normalsize }
    \label{fig:large2}
\end{figure*}


\begin{figure*}
    \centering
    \includegraphics[width=0.45\textwidth]{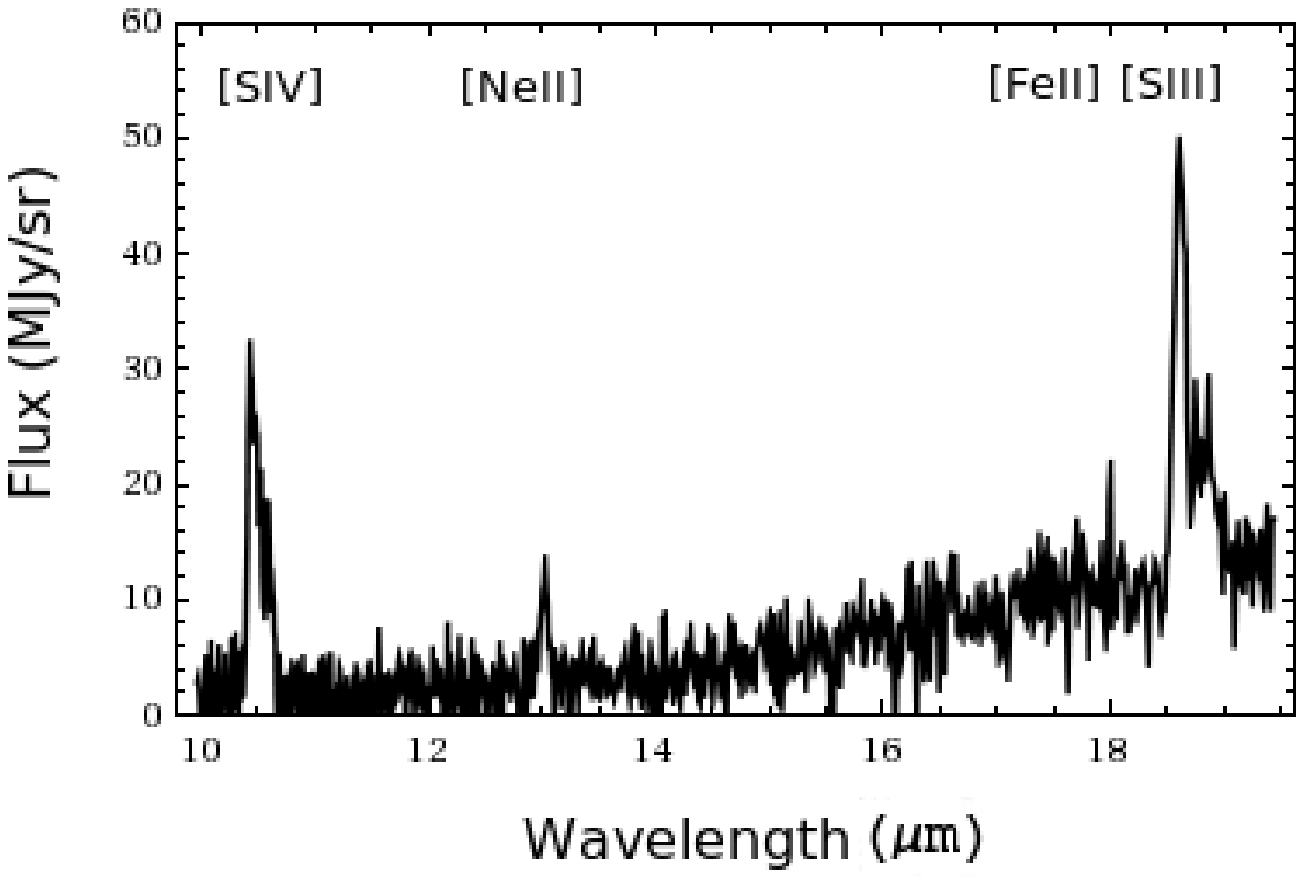}
    \includegraphics[width=0.45\textwidth]{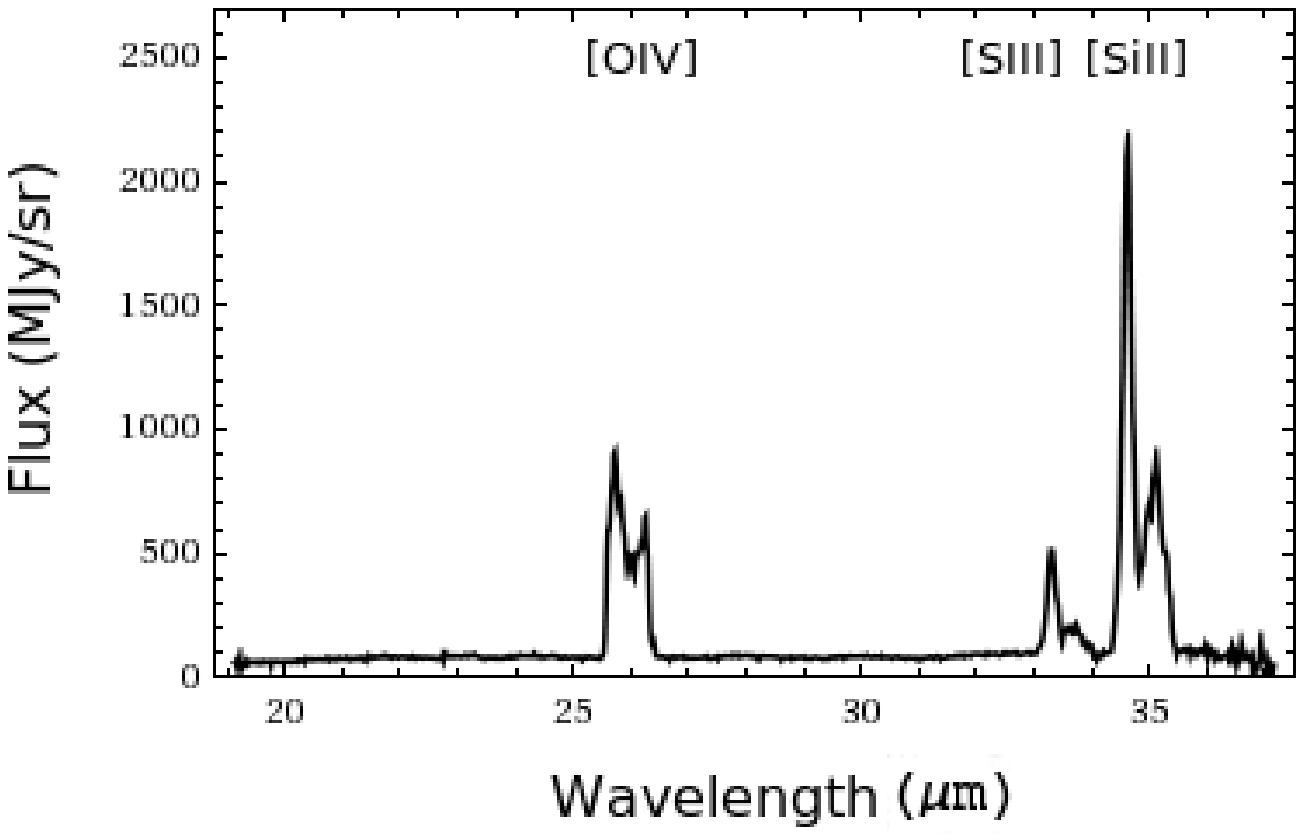}

\caption{\small Typical spectra from the SH and LH \emph{Spitzer} IRS modules of central emission of Cas~A.
\normalsize }
    \label{fig:large3}
\end{figure*}


\begin{figure*}
    \centering
    \includegraphics[width=0.45\textwidth]{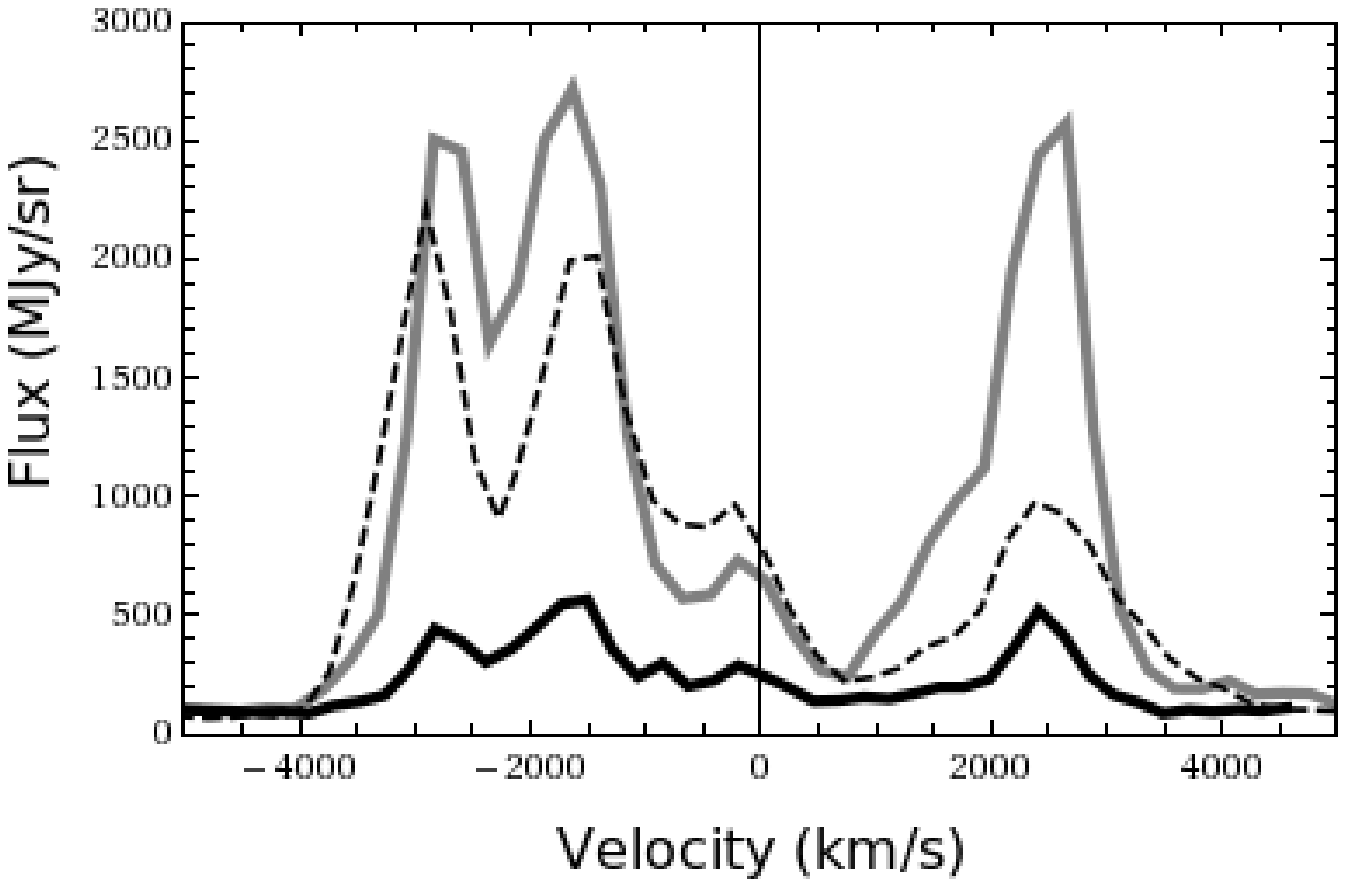}
    \includegraphics[width=0.45\textwidth]{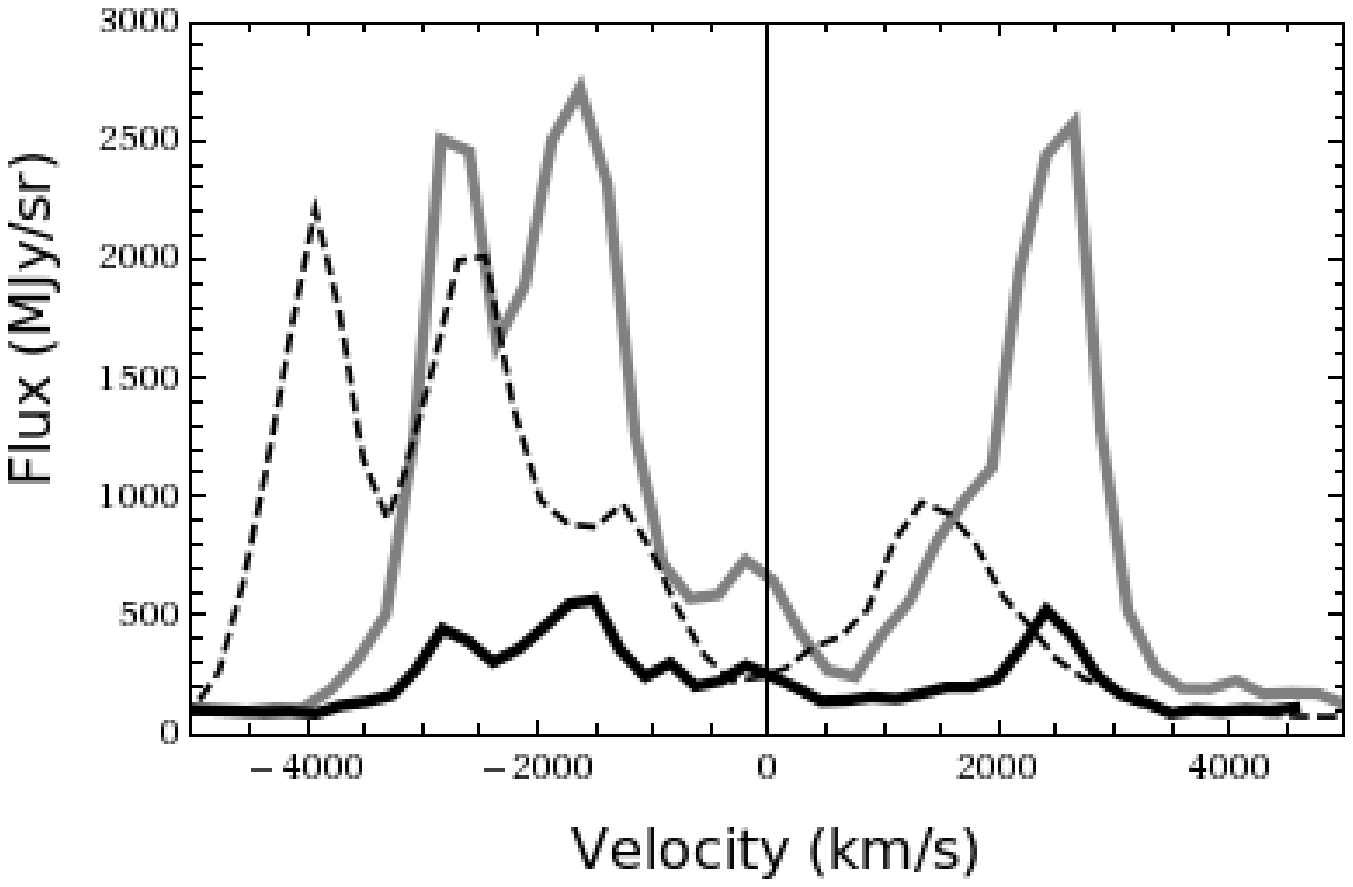}

\caption{\small Velocity plot for the [Si~II] line (gray) and the 33.48$\mu$m [S~III] line (black) over-plotted with the 25.9$\mu$m line (dashed) shifted under the assumption that it is either all [O~IV] (left) or [Fe~II] (right).  The peaks match very well for the assumption that the 25.9$\mu$m line is all O, but match very poorly under the assumption that it is composed of Fe.
\normalsize }
    \label{fig:large4}
\end{figure*}


\begin{figure*}
    \centering
    \includegraphics[width=0.45\textwidth]{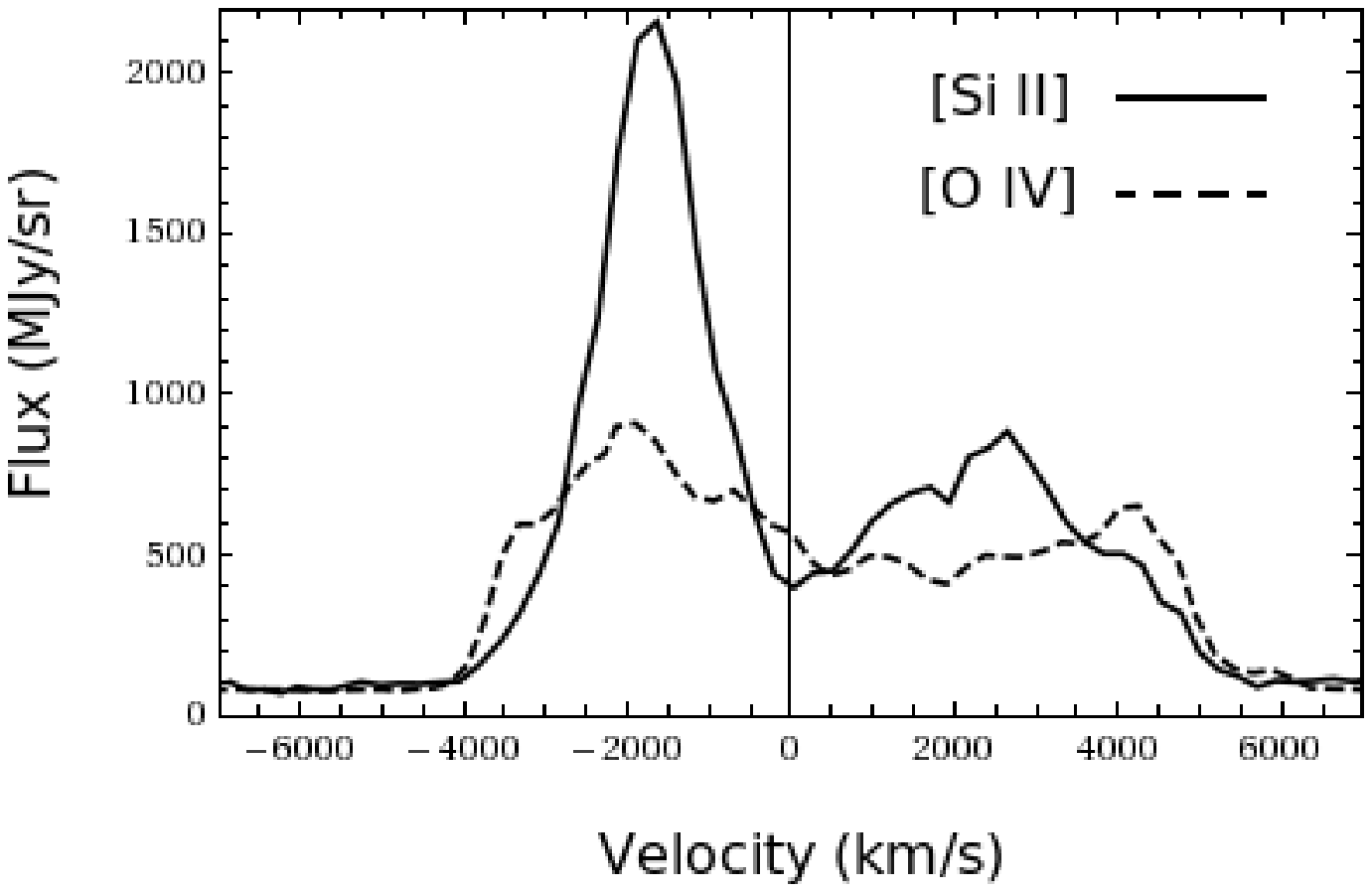}

\caption{\small Doppler structure of [O~IV] and [Si~II] lines integrated over the entire central region.  The average velocity of material on the back of the remnant is $\sim$900 km~s$^{-1}$ greater than that of material on the front.
\normalsize }
    \label{fig:large5}
\end{figure*}


\begin{figure*}
    \centering
    \includegraphics[width=0.9\textwidth]{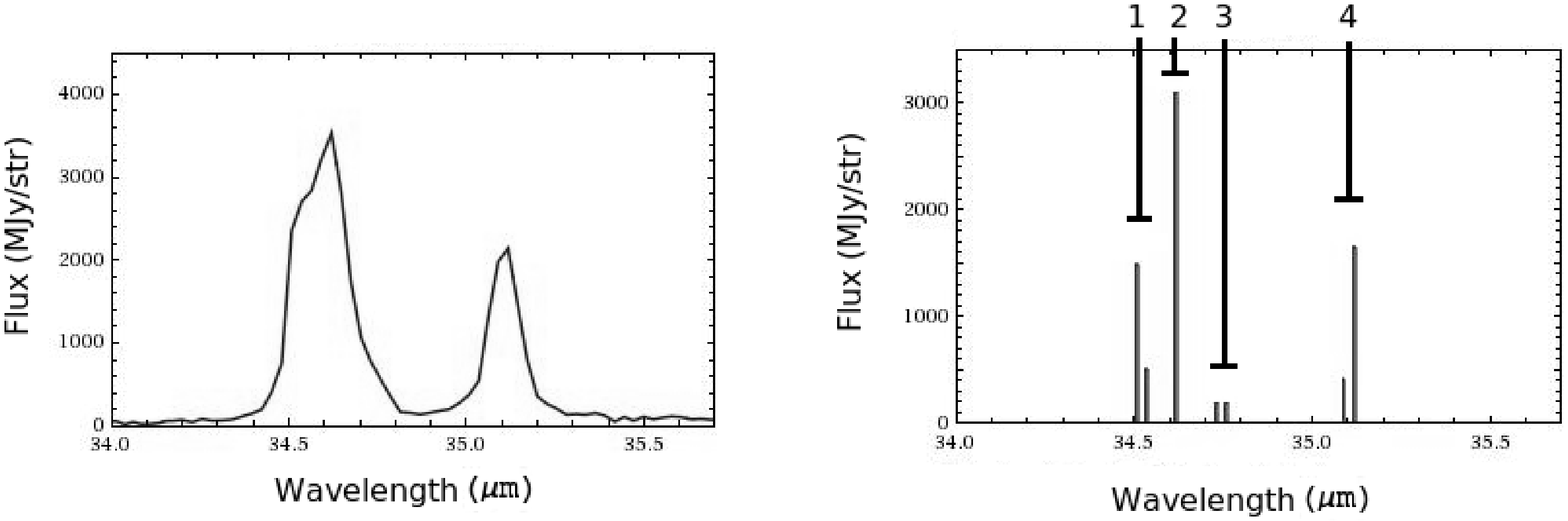}

\caption{\small Spectral CLEAN algorithm applied to a sample spectrum (left). CLEAN components from adjacent bins were combined (as seen in the cases of the neighboring bins at $\sim$34.5$\mu$m, 34.75$\mu$m, and 35.1$\mu$m in the right figure) and very weak components with fluxes less than 100 MJy sr$^{-1}$ were removed, so only 4 distinct Doppler components were extracted from this line of sight.  These components are shown and numbered (right).  The separation between these components is about 850, 1300, and 3000~km~s$^{-1}$ from left to right, comfortably larger than our $\sim$100~km~s$^{-1}$ uncertainty.
\normalsize }
    \label{fig:large6}
\end{figure*}


\begin{figure*}
    \centering
    \includegraphics[width=0.45\textwidth]{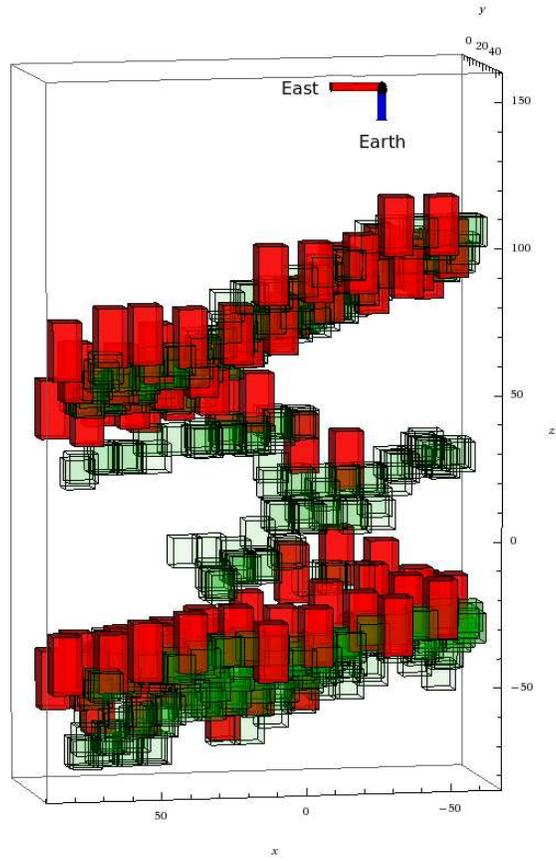}

\caption{\small Comparison of our high resolution [Si~II] line (green) to \cite{delaney09} (red), which is at lower spectral resolution.  The units are arcseconds from the center of the Bright Ring.
\normalsize }
    \label{fig:large7}
\end{figure*}


\begin{figure*}
    \centering
    \includegraphics[width=0.9\textwidth]{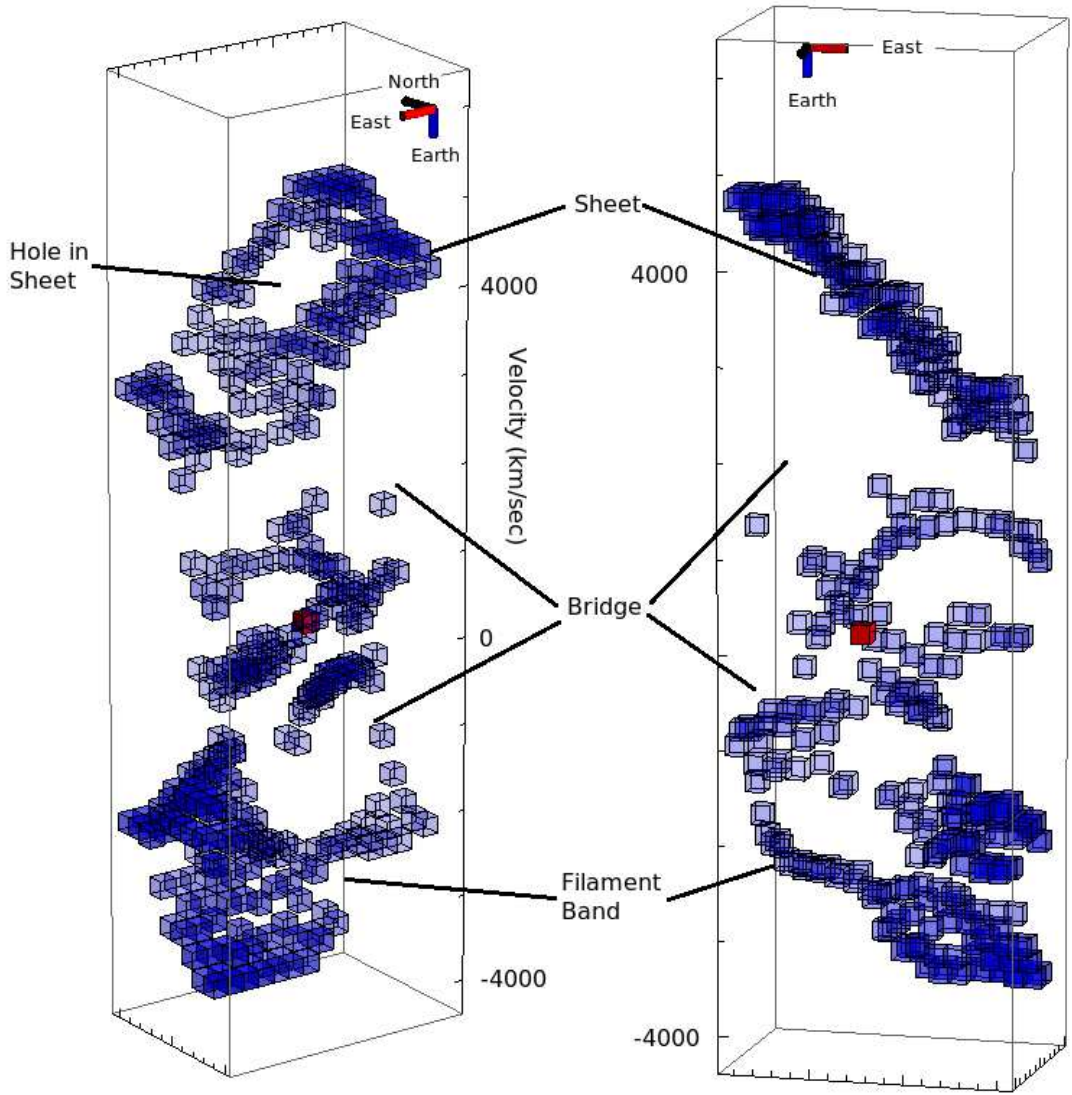}
\caption{\small 3D plot of the the 25.9$\mu$m [O~IV] line (blue) as viewed from two different angles.  The three major structures discussed in the text are labeled.  The red box denotes the location of the central compact object.  The velocity axis has been stretched by a factor of 1.8 to highlight the velocity structures discussed in the text.  The thinness of the Sheet is shown in the figure to the right, where the Sheet's thickness is roughly the minimum thickness allowed by the plotting symbols.
\normalsize }
    \label{fig:large8}
\end{figure*}


\begin{figure*}
    \centering
    \includegraphics[width=0.5\textwidth]{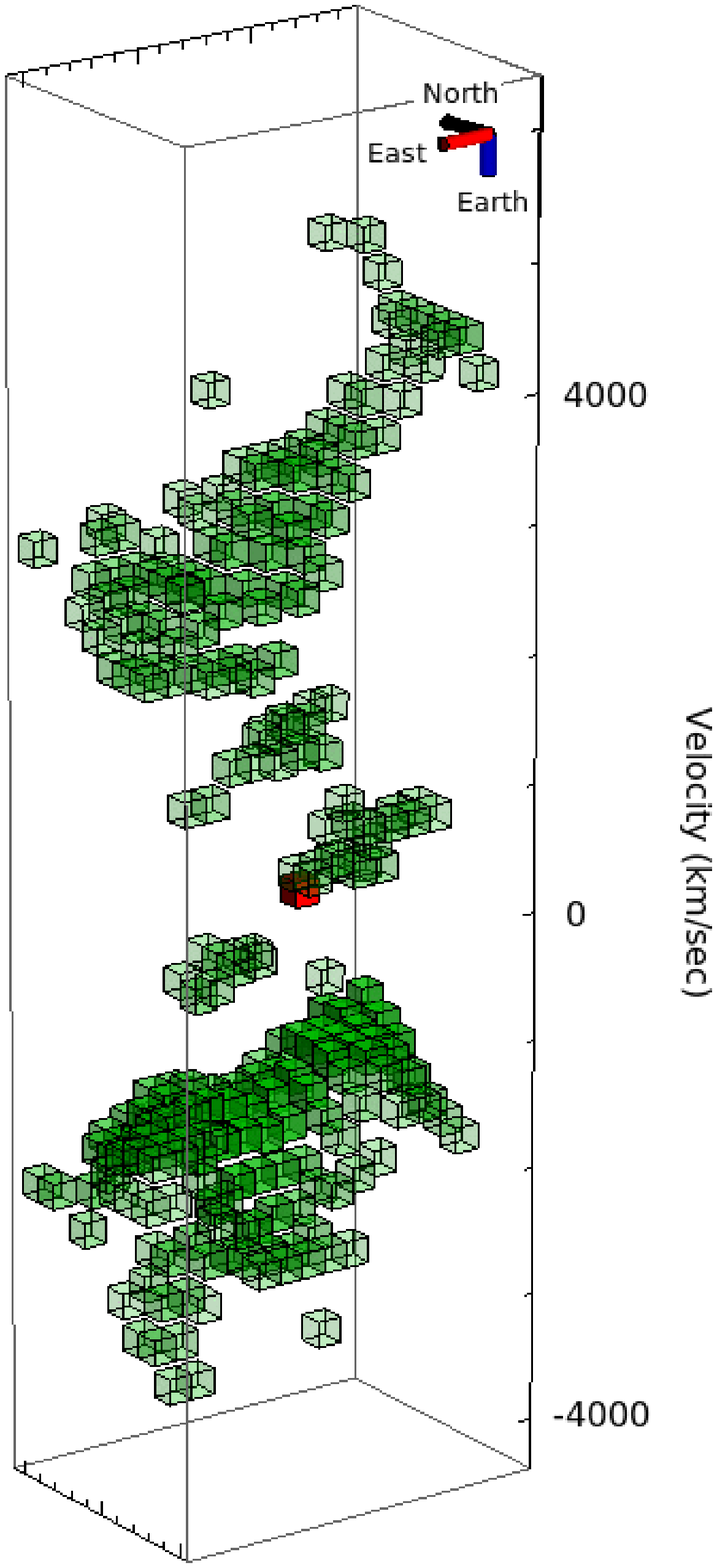}
\caption{\small 3D plot of the 34.81$\mu$m [Si~II] line (green). The red box denotes the location of the central compact object.  The velocity axis has been stretched by a factor of 1.8 to highlight the velocity structures discussed in the text.
\normalsize }
    \label{fig:large9}
\end{figure*}


\begin{figure*}
    \centering
    \includegraphics[width=0.45\textwidth]{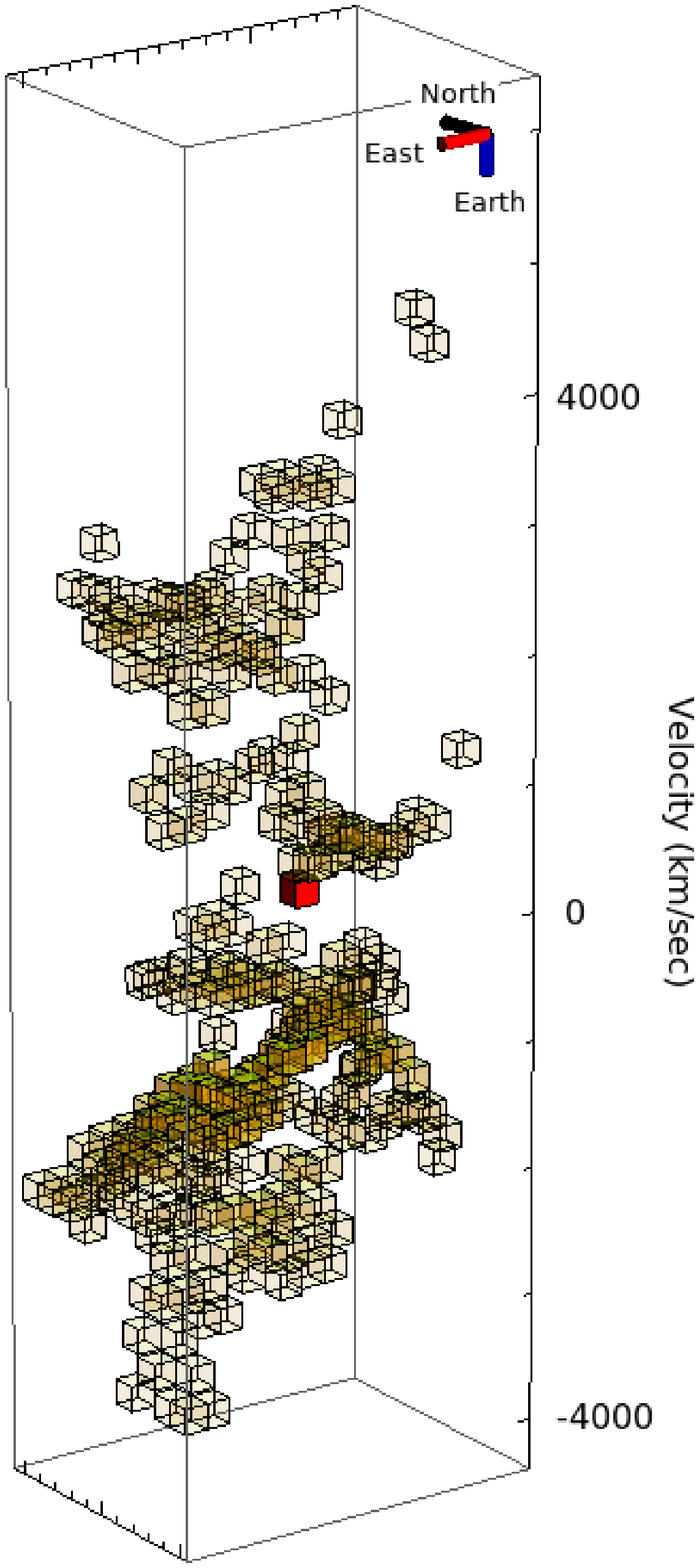}
\caption{\small 3D plot of the 33.48$\mu$m [S~III] line (yellow). The red box denotes the location of the central compact object.  The velocity axis has been stretched by a factor of 1.8 to highlight the velocity structures discussed in the text.
\normalsize }
    \label{fig:large10}
\end{figure*}


\begin{figure*}
    \centering
    \includegraphics[width=0.35\textwidth]{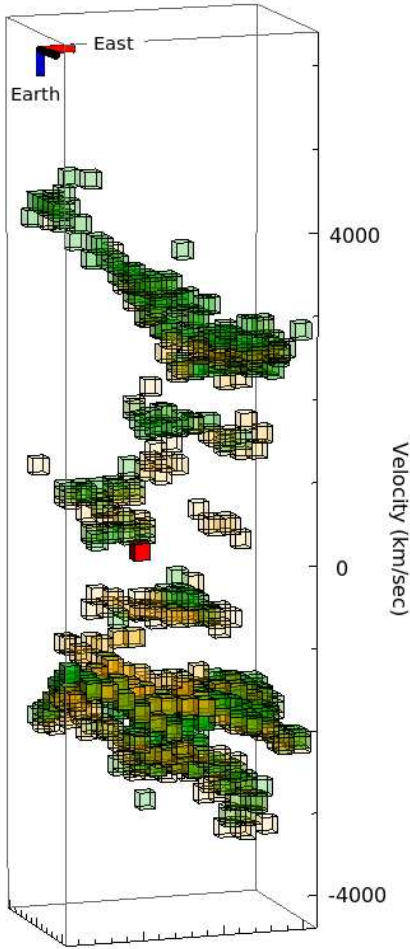}

\caption{\small 3D [Si~II] (green) and [S~III] (yellow) map on the same axes.  The two lines overlap very strongly.
\normalsize }
    \label{fig:large11}
\end{figure*}


\begin{figure*}
    \centering
    \includegraphics[width=0.5\textwidth]{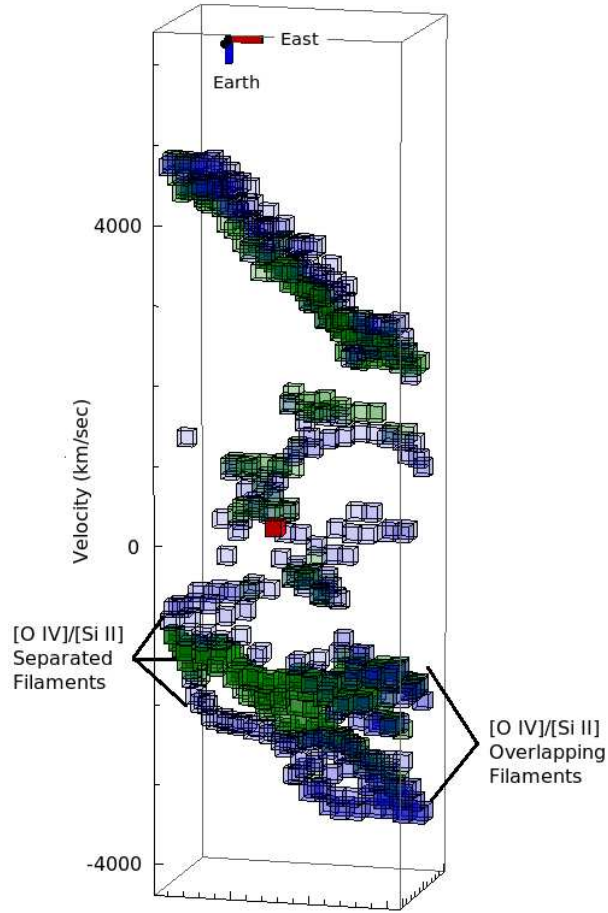}

\caption{\small 3D plot of the 34.81$\mu$m [Si~II] line (green) and the 25.89$\mu$m [O~IV] line (blue) on the same axes.  The red box denotes the location of the central compact object.  The location of the O and Si overlapping and separated filaments are indicated.
\normalsize }
    \label{fig:large12}
\end{figure*}

\begin{figure*}
    \centering
    \includegraphics[width=0.45\textwidth]{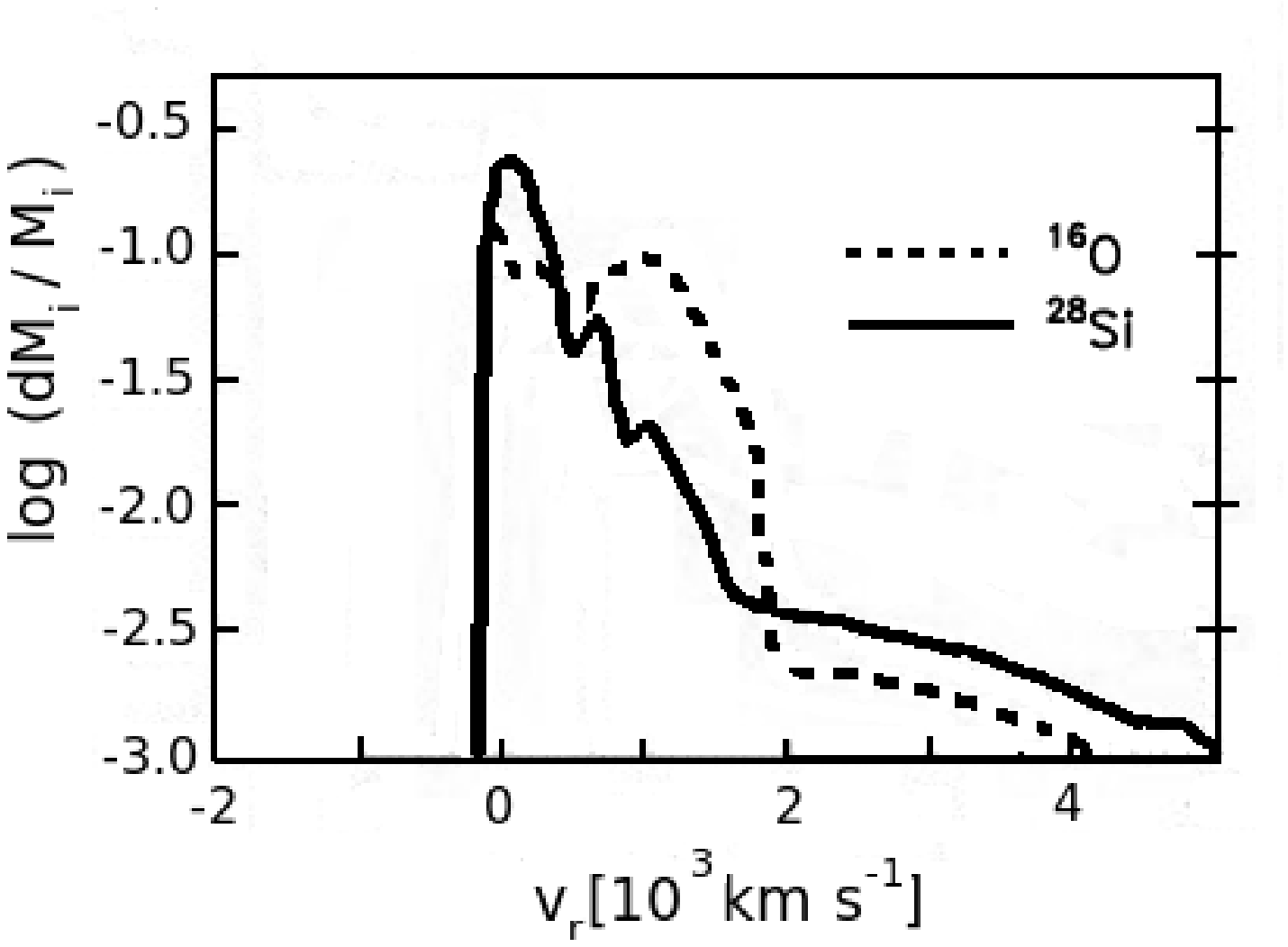}
    \includegraphics[width=0.45\textwidth]{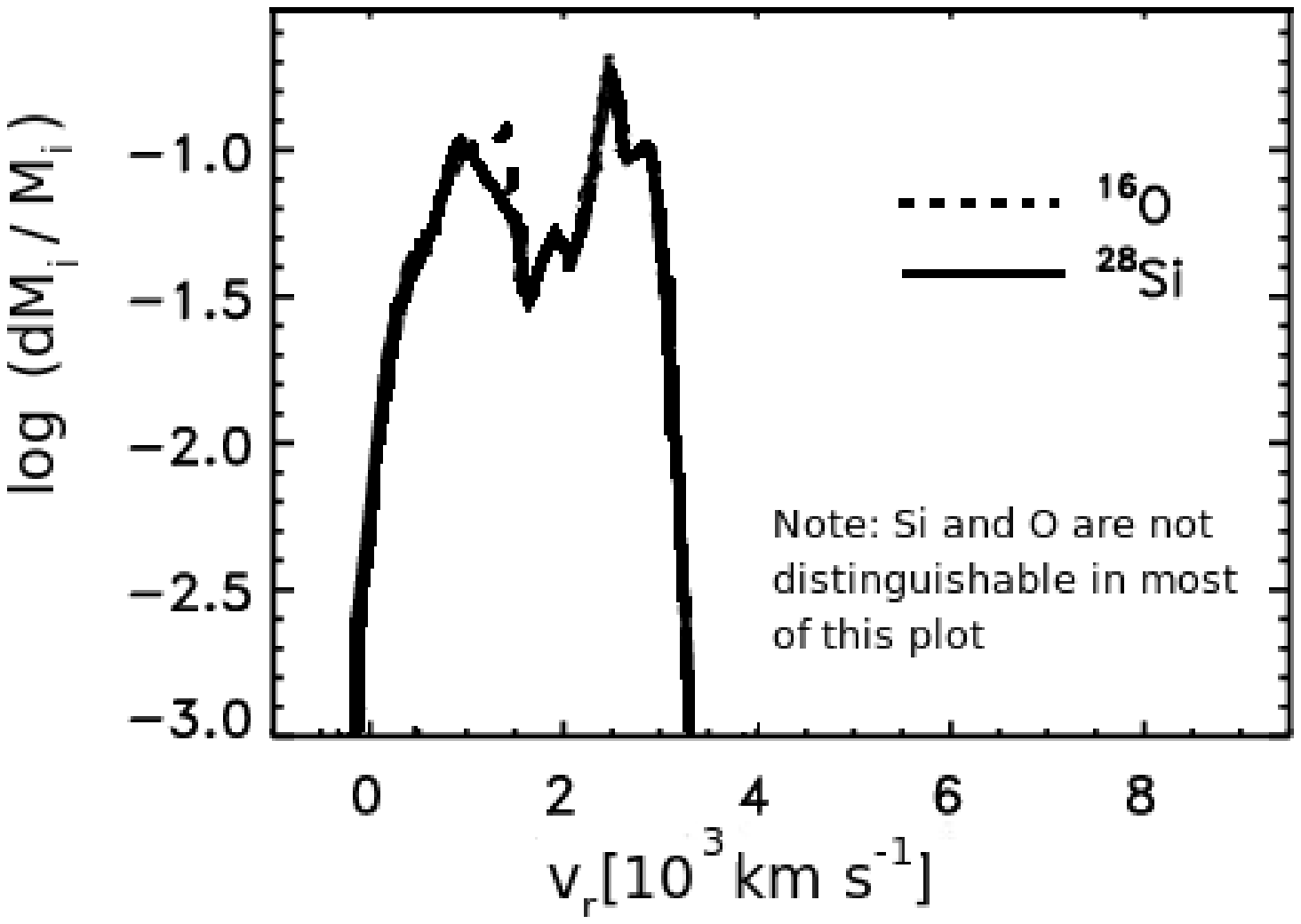}

\caption{\small Radial velocity profiles for two supernova models, at times when the velocity structure has stabilized. Left: 15M$_{\odot}$, solar metallicity star based on \cite{jog09} without rotation. Right: 15M$_{\odot}$, solar metallicity star with rotation based on \cite{kifonidis06}. Si and O layers are shown.  These figures have been altered from their original form for ease of comparison.
\normalsize }
    \label{fig:large13}
\end{figure*}


\begin{figure*}
    \centering
    \includegraphics[width=0.65\textwidth]{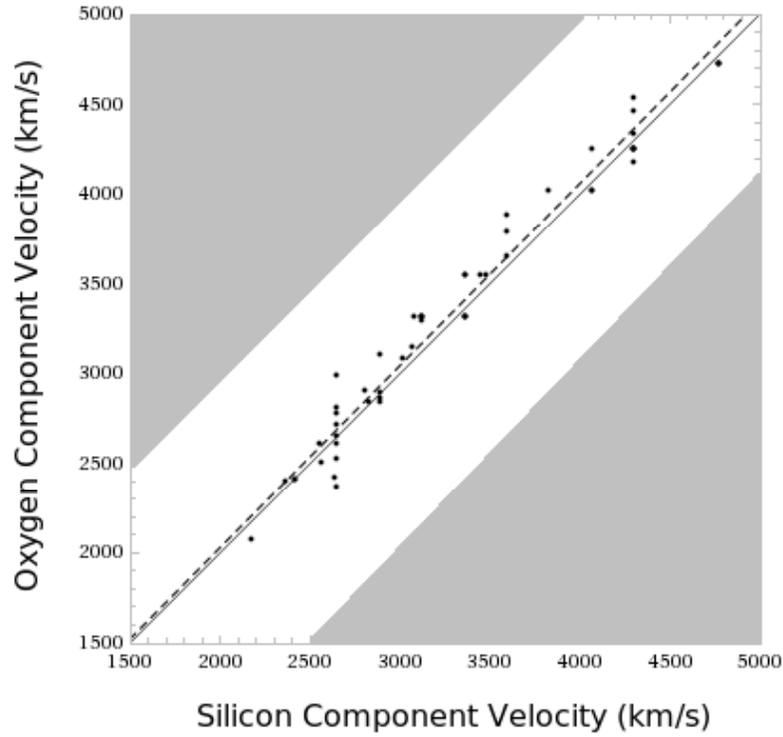}

\caption{\small Si velocity vs O velocity diagram for the Sheet.  The dashed line represents the best fit line to the data, while the solid line is a line with a slope of 1 that passes through the origin.  Error bars on each point are roughly the size of the points used for plotting.  The white region of the figure represents the area in which both components are within 1000 km~s$^{-1}$ of each other.  In principle, points could lie anywhere within the white area and still be considered a Doppler component pair.
\normalsize }
    \label{fig:large14}
\end{figure*}


\begin{figure*}
    \centering
    \includegraphics[width=0.75\textwidth]{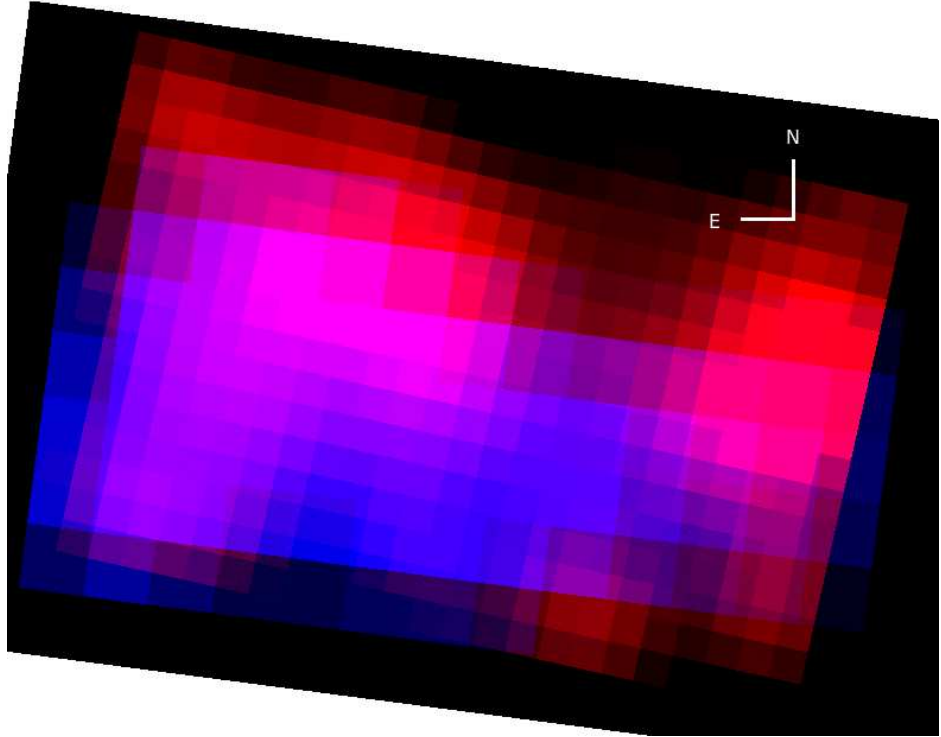}
    \caption{Left: \small Map of [Ne~II] at 12.8$\mu$m (red) and [O IV] at 25.9$\mu$m (blue).  The [Ne~II] map has been smoothed by 2 pixels in order to increase the signal.  There is little correlation between the two ions despite the fact that they originated from the same nucleosynthetic layer.
\normalsize }
    \label{fig:large15}
\end{figure*}


\begin{figure*}
    \centering
    \includegraphics[width=0.75\textwidth]{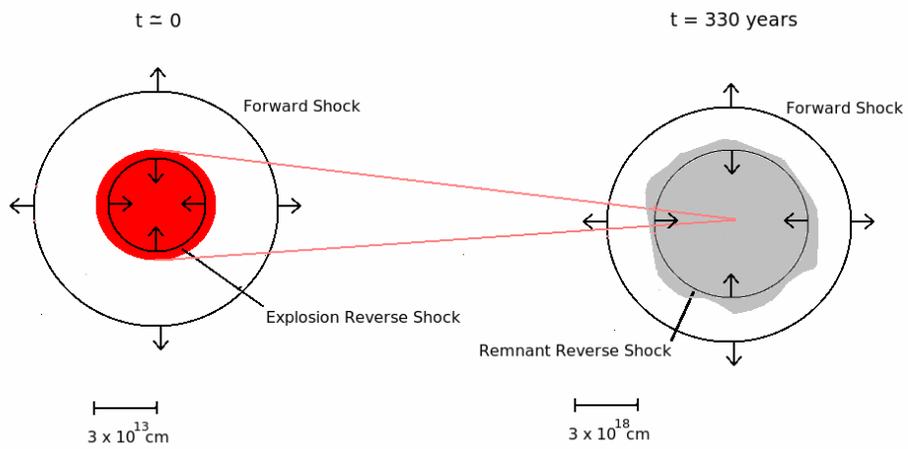}
    \caption{Left: \small The Explosion Reverse Shock and Remnant Reverse Shock with ejecta (gray).  The red circle represents the progenitor star (left).
\normalsize }
    \label{fig:large16}
\end{figure*}

\end{document}